\documentclass[a4paper,11pt]{article}
\pdfoutput=1 

\usepackage{jheppub} 

\usepackage{graphicx}
\usepackage{epstopdf}
\usepackage{amsmath,amssymb,mathrsfs}
\usepackage{ bbold }
\usepackage[dvipsnames]{xcolor}
\usepackage{subcaption}
\usepackage[normalem]{ulem}

\usepackage{tikz}
\usetikzlibrary{decorations.pathmorphing}
\usetikzlibrary{shapes.geometric}
\usetikzlibrary{calc}
\usetikzlibrary{shapes.misc}
\usetikzlibrary{decorations.markings}
\tikzstyle{gluon}=[decorate, decoration={coil,aspect=0.8, amplitude=1.5pt,  segment length=3pt}]
\tikzstyle{lgluon}=[decorate, decoration={coil,aspect=-0.8, amplitude=1.5pt,  segment length=3pt}]
\tikzset{->-/.style={decoration={
  markings,
  mark=at position #1 with {\arrow{>}}},postaction={decorate}}}
\tikzset{-<-/.style={decoration={
markings,
mark=at position #1 with {\arrow{<}}},postaction={decorate}}}

\newcommand{\req}[1]{{Eq.~(\ref{#1})}}

\newcommand{\tr}{\mbox{tr}}
\newcommand{\dd}{\ensuremath{\text{d}^2}}

\newcommand{\thalf}{{\tfrac{1}{2}}}
\newcommand{\hcalf}{{\ensuremath{\hat{\mathcal{F}}}}}
\newcommand{\hcalo}{{\ensuremath{\hat{\mathcal{O}}}}}
\newcommand{\cala}{{\ensuremath{\mathcal{A}}}}
\newcommand{\calb}{{\ensuremath{\mathcal{B}}}}
\newcommand{\calf}{{\ensuremath{\mathcal{F}}}}
\newcommand{\calo}{{\ensuremath{\mathcal{O}}}}

\newcommand{\calw}{{\ensuremath{\mathcal{W}}}}
\newcommand{\calg}{{\ensuremath{\mathcal{G}}}}
\newcommand{\cald}{{\ensuremath{\mathcal{D}}}}
\newcommand{\xvec}{{\ensuremath{\underline{x}}}}
\newcommand{\yvec}{{\ensuremath{\underline{y}}}}

\newcommand{\uvec}{{\ensuremath{\underline{u}}}}
\newcommand{\vvec}{{\ensuremath{\underline{v}}}}
\newcommand{\wvec}{{\ensuremath{\underline{w}}}}
\newcommand{\rvec}{{\ensuremath{\underline{r}}}}

\newcommand{\PointsLines}{{The points correspond to the simulation data, and the lines correspond to the Gaussian truncation. The inputs for the Gaussian truncation are the values of the fundamental dipole's $\Gamma_r^{(ic)}$ and $\gamma_r^{(evo)}$.}}
\newcommand{\PointsLinesAdj}{{The points correspond to the simulation data, and the lines correspond to the Gaussian truncation. The inputs for the Gaussian truncation are the values of the adjoint dipole's $\Gamma_r^{(ic)}$ and $\gamma_r^{(evo)}$.}}

\title{\boldmath Small-\texorpdfstring{$x$}{x} TMD  distributions: JIMWLK evolution and the Gaussian truncation}

\author[a,1]{Florian Cougoulic,\note{Corresponding author.}}
\author[a]{Piotr Korcyl}

\affiliation[a]{Institute of Theoretical Physics, Jagiellonian University, ul. \L ojasiewicza 11, 30-348 Krak\'ow, Poland}

\emailAdd{florian.cougoulic@uj.edu.pl}
\emailAdd{piotr.korcyl@uj.edu.pl}

\abstract{
We extend the systematic study of small-$x$ Transverse Momentum Dependent (TMD) distributions, initiated in \cite{Cougoulic:2026drr}, by including the energy evolution. 
We use the JIMWLK evolution equation to study the energy dependence of three distributions in the quark-gluon sector, $\calf_{qg}$, and seven distributions in the gluon-gluon sector, $\calf_{gg}$.
The initial conditions are provided by the McLerran-Venugopalan model.
We study these TMD distributions in position space and with $N_c = 3$ fixed.
We then compare these results to the Gaussian truncation of the Balitsky hierarchy.
Although the Gaussian truncation worked very well for the TMD distributions at the initial condition, now we identify some cases where the Gaussian truncation provides a reasonably good description, but in the majority of cases it fails to fully reproduce numerical data. 
Using the fixed representation building blocks, $\Omega_{ag}^\omega$ - previously introduced in \cite{Cougoulic:2026drr}, we are able to partially explain the outcomes. 
This work constitutes the next step with the goal of understanding the interplays between the large-$N_c$ limit and Gaussian truncation as well as the BK/JIMWLK evolution equations for phenomenologically relevant observables. 
}

\begin{document} 
\maketitle
\flushbottom

\eject
\section{Introduction}
\label{sec:intro}

Transverse momentum dependent (TMD) distributions provide a framework for describing the transverse-momentum structure of hadrons and play a central role in understanding differential observables in high-energy QCD processes, see \cite{Boussarie:2023izj} for a review. 
In processes characterized by a separation between a hard scale and the observed transverse-momentum scale, the relevant observables can be formulated within a TMD factorization \cite{Collins:2011zzd}.
Over the years, they have been widely used to describe observables in QCD-induced processes, such as: 
distribution of heavy bosons at small transverse momentum 
\cite{Collins:1984kg,Qiu:2000ga,Ji:2004xq},
spin asymmetries 
\cite{Ralston:1979ys,Sivers:1989cc,Sivers:1990fh,Collins:1992kk,Boer:2003cm},
the scattering of hadrons at high-energy 
\cite{Lipatov:1985uk,Catani:1990eg}, and many others.
In recent years, the TMD distributions have also attracted a lot of attention in the context of small-$x$ physics.
In the limit of small Bjorken-$x$, TMD distributions are expressed in terms of operators made of Wilson lines (WL), which are evaluated against a target state described by a \textit{color glass condensate} (CGC). See the reviews \cite{Gelis:2010nm,Albacete:2014fwa,Blaizot:2016qgz,Morreale:2021pnn, Raj:2025hse} and the references therein.
The dependence of the dipole and TMD distributions on rapidity can be obtained using the JIMWLK equation \cite{Jalilian-Marian:1997dw,Jalilian-Marian:1997gr,Weigert:2000gi,Iancu:2001ad,Iancu:2000hn,Ferreiro:2001qy}, alternatively, the Balitsky hierarchy \cite{Balitsky:1995ub,Balitsky:1998kc,Balitsky:1998ya}.
In the case of the dipole, the large-$N_c$ limit yields the BK equation \cite{Balitsky:1995ub,Balitsky:1998ya,Kovchegov:1999yj,Kovchegov:1999ua}.
The small-$x$ evolution resums powers of logarithms of $1/x$, schematically terms of the form $\left[ \alpha_s \ln 1/x \right]^n$, associated with gluon emissions, see for example \cite{Kovchegov:2012mbw}.

While the rapidity evolution given by leading-order JIMWLK evolution applies to any operators, and in particular those appearing in TMD distributions, it has a main technical drawback: solving the JIMWLK equation requires a numerical stochastic method based on the Langevin equation \cite{Blaizot:2002np,Lappi:2012vw}, which is a computationally expensive endeavor. 
On the other hand, solving the BK equation is a more efficient approach. Additionally, this approach has seen significant advancements in recent years.\footnote{
The next-to-leading-order (NLO) BK equation was derived in \cite{Balitsky:2007feb,Balitsky:2013fea}.
Numerical solutions to the NLO BK equation were explored in \cite{Stasto:2013cha,Lappi:2015fma,Ducloue:2016shw}, where negativity and instability issues were found. 
Modifications to the NLO BK have been proposed to address those issues such as  kinematically constrained evolution \cite{Motyka:2009gi,Beuf:2014uia,Iancu:2015vea}, changing the evolution variable from the projectile to the target \cite{Ducloue:2019ezk,Altinoluk:2025tms}, and rotating the basis of the operator in the operator product expansion \cite{Boussarie:2025bpq,Boussarie:2025mzh}.
The non-conformal part of the NNLO BK equation has been derived \cite{Brunello:2025rhh}.}
However, one cannot access the myriad of operators required for the ever-increasing wealth of differential data from modern experiments with only the large-$N_c$ truncation to the Balitsky hierarchy.
A solution to this mismatch is to rely on the Gaussian Truncation (GT) of the Balitsky hierarchy.
Using the GT, any correlation function made of an arbitrary number of WLs is {analytically} expressed in terms of the two-point function $\gamma_r^t$, which in turn is related to the logarithm of the dipole.
The rapidity dependence of the dipole, which can be evaluated according to the computationally much simpler BK equation,  provides the energy dependence of $\gamma_r^t$, which gives a prediction, under the GT,  for the rapidity dependence of any operator one would like to evolve according to the JIMWKL equation.
This approach has been widely used over the years, see {e.g.} \cite{Kovner:2001vi,Iancu:2001md,Iancu:2002xk,Iancu:2002aq,Weigert:2005us,Kovchegov:2008mk,Dumitru:2010ak,Marquet:2010cf,Iancu:2011nj,Lappi:2026kxi} and references therein.
While this program has obvious advantages, it has been shown, {e.g.} \cite{Dumitru:2010ak,Lappi:2026kxi}, that the GT does not automatically capture all relevant features of correlation functions for the quadrupole and the sextupole. 
Thus, it is relevant for phenomenological applications to understand when the GT can be used such that it captures the full features of the complete energy evolution given by the JIMWLK equation.

In this article, we {investigate} the rapidity evolution of the TMD distributions appearing in \cite{Marquet:2016cgx,Bury:2018kvg}, which can be extracted in processes such as di-jet production. 
We also aim to explore the validity of the Gaussian truncation for those distributions.
A novelty here is that we consider the GT of two-point operators, albeit containing a large number of fundamental WLs and their derivatives.
We {probe} those TMD distributions using the ``building block'' operators $\Omega_{ag}^\omega$ identified in \cite{Cougoulic:2026drr}. 
Those operators have a simple interpretation in terms of irreducible representations (irreps), which systematizes the study of the TMD distributions. 
This analysis serves as a baseline at $N_c = 3$ for an upcoming analysis along the lines of \cite{Cougoulic:2026drr}, where we will vary $N_c = \{2,3,4,5\}$ in order to study the interplay between the JIMWLK equation and the BK equation in the context of TMD distributions.

The manuscript is organized as follows.
Section~\ref{sec:unif_description} introduces our notation and the unified description used for the generation of WL realizations according to the MV-model initial condition and evolved according to the JIMWLK equation.
In section~\ref{sec:GT}, we introduce the Gaussian truncation to the Balitsky hierarchy and extract, from the simulation data of the JIMWLK equation, the dipole evolution and the corresponding two-point correlation $\gamma_r^t$ to be used in all evaluation of the TMD distribution in the GT.
Section~\ref{sec:Omega} recalls the operators $\Omega_{ag}^\omega$, and we compare their prediction from the GT to the simulation data.
In section~\ref{sec:tmd}, we use the previously defined $\Omega_{ag}^\omega$ to evaluate TMD distributions and illustrate the main features of the GT compared to the simulation data.
Finally, section~\ref{sec:discussion} is dedicated to the discussion and prospects of this project.

\section{MV initial condition and JIMWLK evolution - unified description}
\label{sec:unif_description}

In this section, we introduce our formulation.
Due to the similarities between the Langevin formulation \cite{Blaizot:2002np,Lappi:2012vw} of the JIMWLK evolution 
\cite{Jalilian-Marian:1997dw,Jalilian-Marian:1997gr,Weigert:2000gi,Iancu:2001ad,Iancu:2000hn,Ferreiro:2001qy} and the construction of initial conditions according to the MV-model \cite{McLerran:1993ni,McLerran:1993ka,McLerran:1994vd}, we unify the notation to establish a connection with the building blocks of the Gaussian truncation.
We start by discussing an example of a single Wilson Line (WL) in section \ref{sec. WL}, before discussing the complete setup over the transverse plane in \ref{sec. workflow}.

\label{subsec:JIMWLK}

\subsection{Building a Wilson line}
\label{sec. WL}

Fundamental Wilson lines are elements of $SU(N_c)$ and can therefore be generated using the Lie algebra.
Starting from the identity, any large gauge transformation in $SU(N_c)$ can be recast as the product of a large number of small (infinitesimal) rotations.
After performing $n$ small rotations, one \textit{updates} a Wilson line $V^{(n)}$ according to:
\begin{equation}\label{eq:wl_ic}
    V^{(n)} \longrightarrow V^{(n+1)} =  e^{i t^a \theta^a_{n+1}} \ V^{(n)}
\end{equation}
where $t^a$ are the fundamental generators with $a \in \{ 1, \cdots N_c^2-1\}$, and $\theta^a$ is a random vector to be specified later for the MV-model.
For the purposes of this study, we assume that the initial condition is independent of the longitudinal profile of the target.
Thus, all random variables $\theta_{n}$ are distributed according to the same probability distribution. 
At the initial condition level, it is irrelevant whether the small rotation, in \req{eq:wl_ic}, acts to the left or right of the WL.

This feature is not preserved when the JIMWLK evolution is performed.
According to the Langevin formulation of the leading-order JIMWLK \cite{Blaizot:2002np,Lappi:2012vw}, one generates a random variable on one side of the shock wave and relates it to the other side by multiplying it by an adjoining Wilson line. 
This implies a distinction between left and right multiplication that must be taken into account.
The rapidity step, for a WL already evolved at the step $m$, reads formally
\begin{equation}
    V^{(m)} \longrightarrow V^{(m+1)} = e^{+ it^a \eta^a_{m+1,>}} V^{(m)} e^{- it^b \eta^b_{m+1,<}}
\end{equation}
where $\eta^a_{\lessgtr}$ denotes the random right/left rotation angle used to \textit{update} the WL, and will be specified later for the action of the JIMWLK Hamiltonian. This formulation guarantees that the result belongs to $SU(N_c)$  at any step $m$.

Let us denote the number of rotations used to generate the initial condition for the Wilson line by $N \in \mathbb{N}$, and the number of steps in rapidity evolution within a rapidity interval\footnote{The rapidity $Y$ and the coupling $\alpha_s$ only appear as a product $\alpha_s Y$ in the leading-order JIMWLK evolution. Thus, we can simply trace $Y$ and set $\alpha_s =1$ to streamline the notation, i.e., each occurrence of $Y$ should be understood as $\alpha_s Y$ when it is not explicitly written as such.} of interest $Y$ by $M \in \mathbb{N}$. 
A Wilson line, evolved to this scale $Y$ using this discretization, takes the form
\begin{equation}\label{eq:wl_constr}
    V^{[N,M]} =
    e^{+ i \eta_{M,>}} e^{+ i \eta_{M-1,>}} \cdots e^{+ i \eta_{1,>}}  
    \left(
    \prod_{n=1}^N e^{i \theta_n}
    \right)
    e^{- i \eta_{1,<}} \cdots e^{- i \eta_{M-1,<}} e^{- i \eta_{M,<}}
\end{equation}
where we use the shorthand notation $\eta_{\lessgtr} = t^a \eta^a_{\lessgtr}$, and $\theta = t^a \theta^a$. 
For a fixed $Y$, as $N$ and $M$ become large, one recognizes the expression for the discretization of a path-order integral: 
\begin{align}\label{eq:wl_path-order_rapidity}
    V^Y = P \exp \left[ i\, t^a \int\limits_{-(1+Y)}^{1+Y} dt\ \alpha^a(t)\right]
\end{align}
The argument $t$ in $\alpha(t)$ of \req{eq:wl_path-order_rapidity}  does not denote a light-cone time. It simply orders the rotations due to the generation of the initial condition and the rotations due to the evolution. 
The explicit form of $\alpha(t)$ either in the initial condition, i.e. $\theta$ for $|t| \leq 1$, or during the evolutions, i.e. $\eta_{\lessgtr}$ for $|t| > 1$, is shown below in \req{eq:ic_alpha} and \req{eq:evo_alpha}.

Let us note the analogy of this notation with the usual definition of a Wilson line along the light-cone direction $x^\mu_{proj} = (n\cdot x)\,\bar{n}^\mu$, with $n^2 = \bar{n}^2 = 0$ and $n \cdot \bar{n} =1$. 
In the projectile light-cone gauge $A \cdot n = 0$, we denote by $\alpha$ the large {eikonal} component of the gauge field such that
\begin{equation}
    A^\mu dx_\mu \sim (A\cdot \overline{n})\ d(n\cdot x) \equiv \alpha\ dt
\end{equation}
where $t$ is the corresponding light-cone time $t = n \cdot x$. A WL at the transverse position $\xvec$ between the time $t_1<t_2$ reads
\begin{equation}
    V_\xvec[t_2,t_1] = P \exp \left[ ig\,t^a \int\limits_{t_1}^{t_2}dt\ \alpha^a(t,\xvec) \right].
\end{equation}
In the CGC approach, the target's support is assumed to be Lorentz-contracted as the center-of-mass energy $s$ is taken to infinity. 
The renormalization equation for energy dependence requires us to integrate quantum fluctuations successively around the initial support of the target.
After integrating each fluctuation, the result can be interpreted as the target's support extending along the light cone \cite{Iancu:2011nj}.
In this picture, it is reasonable to link the rapidity variable $y\in (0,Y)$ to the growing support of the target $|t|$. 
A slight distinction compared to the literature, e.g. \cite{Lappi:2026kxi}, is that we encode a support for the initial condition $|t| \leq 1$ to make the link with our numerical implementation concrete.
This support is convenient to study TMD distributions since the derivatives in the operator definition (see the appendix \ref{App:operators}) can act on fields $\alpha(t)$ belonging to the initial condition, {i.e.} $|t| \leq 1$.

\subsection{Workflow}
\label{sec. workflow}

We briefly describe the necessary steps to numerically evaluate any observable made of WLs within the CGC-JIMWLK approach.

\subsubsection{Geometry considerations}

We assign a fundamental Wilson line to each position $\xvec$ in the plane perpendicular to the two light-cone directions $n^\mu$ / $\bar{n}^\mu$.
To minimize the impact of infrared effects on our results, we impose a periodic boundary condition that transforms the full Euclidean transverse plane into a torus with dimensions $L \times L$.
The scale $L$ is chosen such that results depend weakly on the infrared parameter within a few units of the saturation radius $r_{sat} \sim 1/Q_s$. 
In practice, a difference of orders of magnitude is required between the length of the torus $L$ and the scale of the problem $r_{sat}$, in order to evaluate the initial condition for evolution \cite{Korcyl:2021pef}.
Since we are interested in TMD distributions expressed from the derivatives of WLs, this large ratio $L/r_{sat}$ is problematic for discretizing the torus because it automatically requires a very large number of grid points.
As we will see below, we introduce an additional infrared regulator $m$ at the level of the Green's function for both the initial condition and the evolution kernel (for consistency). 
This infrared scale can be adjusted to be closer to the scale of interest. This reduces the sensitivity of our results to the large infrared scale  $L$, enabling us to consider smaller values of the $L/r_{sat}$ ratio. Consequently, we can use a smaller lattice spacing, which is useful for evaluating derivatives with finite differences, particularly after evolution, when distributions drift toward the ultraviolet (UV).
Given this geometry, we define distances between two points as the shortest length on the torus.

\subsubsection{Initial condition}
The only parameter of the MV-model is the transverse color charge density $\mu^2$. 
We define a large number $N$, and let us introduce the color charge density for each transverse sheet according to the following:
\begin{equation}
    \widetilde{\mu}^2 = \mu^2/N.
\end{equation}
Under the assumption of the MV-model, color sources $\rho$ are \underline{locally} correlated, and we write
\begin{equation}
    \langle \rho_n(\uvec) \rho_m(\vvec) \rangle = \widetilde{\mu}^2 \delta^{nm} \delta^2(\uvec-\vvec) \overset{discr.}{\longrightarrow} \frac{\widetilde{\mu}^2}{a^2} \delta^{nm} \delta^2_{\hat{\uvec}\,\hat{\vvec}}
\end{equation}
where the label $n$ of the source $\rho_n$ denotes the transverse sheet. The hat on the transverse coordinates denotes the integer position on the transverse lattice, whose spacing is given by $a$ in both directions.
The field is obtained using the Green function $\calg_m$ for the massive Laplace equation 
\begin{equation}\label{eq:ic_alpha}
    \alpha_n(\xvec) = \int \dd\uvec\ \calg_m(\xvec-\uvec) \rho_n(\uvec)
\end{equation}
with:
\begin{equation}
    \lim\limits_{m \rightarrow 0 } \calg_m(\rvec) = \lim\limits_{m \rightarrow 0 } K_0(m |\rvec|) = - \ln \Lambda |\rvec|
\end{equation}
where $\Lambda$ is an infrared regulator.
We identify the field $\alpha_n$ with the angle $\theta_n$ within \req{eq:wl_constr}.

\subsubsection{Evolution}

One chooses either side of the shockwave by convention and generates a two-dimensional noise $\xi^j_m(\wvec)$ according to a normal distribution on the whole torus (the transverse label is $j =\{1,2\}$  and $m \in \mathbb{N}$ labels the rapidity step). 
This noise can be expressed on the other side of the shockwave owing to
\begin{equation}
    \xi^b_{m,>}(\wvec)  = U^{ba}_\wvec \xi^a_{m,<}(\wvec).
\end{equation}
The symbol $U_\wvec$ denotes an adjoint Wilson line implicitly built out of the previously constructed fields $\alpha_t$ in the support $t \in \left( -|y_{m-1}+1|, |y_{m-1}+1|\right)$ with $y_{m-1}$ the rapidity corresponding to the previous step $m-1$.

\noindent
The field at the step $m$ is obtained according to
\begin{equation}\label{eq:evo_alpha}
    \alpha_{m,\lessgtr}(\xvec) = \frac{\sqrt{\delta_y}}{\pi} \int \dd\wvec\ \frac{(\wvec-\xvec)^j}{(\wvec-\xvec)^2} \xi_{m,\lessgtr}^j(\wvec)
\end{equation}
where $\delta_y = Y/M$.
Then we identify the field $\alpha_{m,\lessgtr}$ with the angle $\eta_{m,\lessgtr}$ within \req{eq:wl_constr}.
The left and right rotations generating the energy-evolution take the form\footnote{{To establish a connection with the path-ordered integral, it is helpful to rescale the auxiliary background, denoted by $\xi$, such that the expectation value of two of these noises scales with $1/\delta_y$.}}
\begin{equation}
    e^{\mp i \eta_{m,\lessgtr}} = \exp \left[ \mp i\, t^a \frac{\sqrt{\alpha_s Y/M}}{\pi} \int \dd\wvec\ \frac{(\wvec-\xvec)^j}{(\wvec-\xvec)^2} \xi_{m,\lessgtr}^{j,a}(\wvec)\right].
\end{equation}
For consistency, one can remark that
\begin{equation}
    \lim\limits_{m \rightarrow 0} \nabla^j_\wvec\ \calg_m(\wvec-\xvec) = \frac{(\wvec-\xvec)^j}{(\wvec-\xvec)^2}
\end{equation}
and regularize the JIMWLK kernel in the IR according to the same regularization as the initial condition MV-model. As we evaluate gauge-invariant quantities, this infrared regularization is harmless for the evolution.

Both \req{eq:ic_alpha} and \req{eq:evo_alpha} involve a convolution in the transverse plane, which can be sped up using Fourier transforms \cite{Rummukainen:2003ns}.
We perform the product between the Green function $\widetilde{\mathcal{G}}_m$ and the noise $\widetilde{\xi}$ directly generated in momentum space\footnote{The tilde denotes the Fourier-transformed quantity, i.e., in momentum space.}, then the field $\alpha(\xvec)$ is recovered by performing a two-dimensional fast Fourier transform (FFT) back to position space.

\subsubsection{Operator expectation value}

Let us denote by $\hcalo$ an arbitrary operator made of several Wilson lines at some transverse positions (which can be shared between some of them). We denote by the angle bracket the formal CGC average as defined by the functional integral weighted by the functional $\calw_Y$ at the rapidity $Y$. In numerical implementation, it is more convenient to talk about ensemble average.
We stochastically generate realizations $r$ for the WLs $V^{r,Y}$ at a given rapidity $Y$, from which we construct the operator of interest $\calo^{r,Y}$. 
Then, we average over a large number of such realizations to recover the CGC average:
\begin{equation}
    \underbrace{\langle \hcalo \rangle_Y}_{\text{CGC average}} = \underbrace{\int \cald \alpha\ \hcalo[\alpha]\ \calw_Y[\alpha]}_{\text{functional definition}} = 
    \underbrace{\lim\limits_{Q \rightarrow \infty}\frac{1}{Q}\sum_{r=1}^Q\calo^{r,Y}}_{\text{Ensemble average}}.
\end{equation}

\section{The Gaussian truncation}
\label{sec:GT}

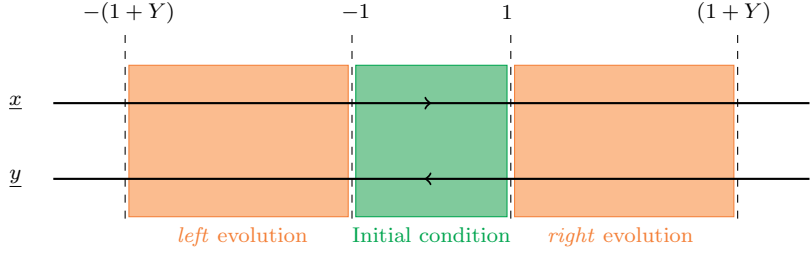
\begin{figure}
    \centering
    \begin{tikzpicture}
    \draw[Green,fill=Green!50!white] (-1,-1) rectangle (1,1);
    \node[below] at (0,-1) {\scriptsize \textcolor{Green}{Initial condition}};

    \draw[Orange,fill=Orange!50!white] (1.1,1) rectangle (4,-1);
    \node[below] at (2.5,-1) {\scriptsize \textcolor{Orange}{\textit{right} evolution}};

    \draw[Orange,fill=Orange!50!white] (-1.1,1) rectangle (-4,-1);
    \node[below] at (-2.5,-1) {\scriptsize \textcolor{Orange}{\textit{left} evolution}};

    
    \draw[thick,->-=.5] (-5,.5) -- (5,.5);
    \draw[thick,-<-=.5] (-5,-.5) -- (5,-.5);

    \node at (1,1.7) {\scriptsize $1$};
    \node at (-1,1.7) {\scriptsize $-1$};
    \node at (4,1.7) {\scriptsize $(1+Y)$};
    \node at (-4,1.7) {\scriptsize $-(1+Y)$};

    \draw[dashed] (1.05,1.4) -- ++(0,-2.5);
    \draw[dashed] (-1.05,1.4) -- ++(0,-2.5);
    \draw[dashed] (4.05,1.4) -- ++(0,-2.5);
    \draw[dashed] (-4.05,1.4) -- ++(0,-2.5);

    \node at (-5.5,.5) {\scriptsize $\xvec$};
    \node at (-5.5,-.5) {\scriptsize $\yvec$};
\end{tikzpicture}
    \caption{Region of interest for the evaluation of a fundamental dipole $\langle \hat{S}_{\xvec\,\yvec} \rangle_Y$ using the symmetric Gaussian truncation.
    The external orange areas represent the support for the evolution in the Gaussian truncation. The inner green area represents the support for the initial condition according to the MV-model. The horizontal lines are Wilson lines (path-ordered exponential) which are evaluated according to the Gaussian approximation involving the interaction potential $\gamma_{\uvec,\vvec}^t$.}
    \label{fig:GT_dipole}
\end{figure}

The keystone in the Gaussian truncation is the two-point correlation function that we denote by $\overline{\gamma}^t_{\uvec - \vvec}$. We introduce it in the subsequent section \ref{sec. 2pt}, then show its relation to the dipole TMD in section \ref{sec. dipoles}. We use this relation in Sections \ref{sec. numerics} and \ref{sec. fits} to extract $\overline{\gamma}^t_{\uvec - \vvec}$ from numerical data: first, we provide the details of our simulations; then, we describe the parametrization that we used to model $\overline{\gamma}^t_{\uvec - \vvec}$. The fitted values of the parameters are gathered in Appendix \ref{App:fit}. We discuss the Casimir scaling violation between the dipole in the fundamental and adjoint representations in section \ref{sec. scaling violation}.

\subsection{Two-point correlation function}
\label{sec. 2pt}

We define $\overline{\gamma}^t_{\uvec - \vvec}$ such that we have:
\begin{equation}\label{eq:fields_corr}
    \left\langle \alpha^a(t,\uvec) \alpha^b(t',\vvec) \right\rangle = \delta^{ab} \delta(t-t')\ \overline{\gamma}^{t}_{\uvec -\vvec}.
\end{equation}
where, in the {r.h.s.}, we emphasize the translation invariance of the setup by writing the argument $\uvec - \vvec$.
Note that for $|t| \leq 1$ the locality in time is one of the assumptions of the MV-model; in the longitudinal direction, each slice of nuclear medium is uncorrelated with the others.
On the contrary, for $|t| > 1$, which corresponds to the new fields generated from the energy evolution, this form is a priori unspecified. The Gaussian truncation \cite{Kovner:2001vi,Iancu:2001md,Iancu:2002xk,Iancu:2002aq,Weigert:2005us,Kovchegov:2008mk,Dumitru:2010ak,Marquet:2010cf,Iancu:2011nj,Lappi:2026kxi} of the Balitsky hierarchy corresponds to assuming the form \req{eq:fields_corr} for the two-point function.
The functional form of $\overline{\gamma}_\rvec^t$ is {numerically} obtained from the evolution of the dipole amplitude, which can be performed according to the BK evolution \cite{Balitsky:1995ub,Balitsky:1998ya,Kovchegov:1999yj,Kovchegov:1999ua}. 
Any other expectation value of an even number of fields $\alpha$ is obtained from this two-point expectation value through the use of Wick contractions, due to the Gaussian ansatz.

While $\overline{\gamma}_\rvec^t$ has a clear interpretation in terms of a two-point expectation value, it will be convenient to work instead with the linear combination (the factor $2$ is reminiscent of \cite{Kovner:2001vi})
\begin{equation}\label{eq:gamma}
    \gamma_\rvec^t = {2} \left( \overline{\gamma}_\rvec^t - \overline{\gamma}_0^t \right),
\end{equation}
which appears when one is interested in evaluating (gauge-invariant) contractions of several WLs.
This follows from the fact that we now assume all correlations to be local in time; one simply uses color conservation to write 
\begin{equation}
    t^a_i t^a_i = t^a_i \sum_{k\neq i} (-t^a_k),
\end{equation} 
where $i$ and $k$ label partons which belong to the gauge-invariant quantity of interest, e.g. $\tr V_1 V_2^\dagger \cdots V_3 V_4^\dagger$. 
This simplifies the double sum over all partons, yielding the factor $2$ in \req{eq:gamma}.
This combination is advantageous as one simply has to consider the sum over distinct pairings: the tadpole contributions are automatically included.
Furthermore, we automatically recover color transparency in the limit $\rvec \rightarrow 0$ from the definition of $\gamma_\rvec^t$.

\subsection{Dipoles}
\label{sec. dipoles}

We define a dipole operator in the irrep $R$ to be
\begin{equation}
    S^{R}_{\xvec\yvec} = \frac{1}{K_R} \tr\, W^{(R)}_\xvec  W^{(R)\dagger}_\yvec,
    \qquad K_R = \tr (\mathbb{1}_R) = \tr (W^{(R)}_\xvec  W^{(R)\dagger}_\xvec).
\end{equation}
The expectation value of this dipole at the rapidity $Y$ will be denoted by
\begin{equation}
    S^{R,Y}_{\xvec\yvec} = \frac{1}{K_R} \left\langle \tr\, W^{(R)}_\xvec W^{(R)\dagger}_\yvec \right\rangle_Y .
\end{equation}
Using the Gaussian truncation, increasing the rapidity by an infinitesimal step $Y \longrightarrow Y + \delta Y$ yields:
\begin{equation}
     S^{R,Y+\delta Y}_{\xvec\yvec} = S^{R,Y}_{\xvec\yvec} - C_R\, {\delta Y} \left( \gamma_{\xvec-\yvec}^{1+Y} + \gamma_{\xvec-\yvec}^{-(1+Y)} \right)S^{R,Y}_{\xvec\yvec}.
\end{equation}
This can be iterated to obtain a solution for the Gaussian truncation of the dipole
\begin{equation}
    S_{\xvec\yvec}^{R,Y} = \exp \left\{ -C_R \int\limits_{-(1+Y)}^{1+Y} dt\,  \gamma^t_{\xvec-\yvec} \right\}.
\end{equation}
{Figure~\ref{fig:GT_dipole}} shows the support of the fields used in the Gaussian truncation to evaluate the fundamental dipole. 
The only difference in this approach compared to evaluating correlators in the MV-model is that we now work with a time-dependent interaction potential, denoted by $\gamma_\rvec^t$. 
In the green shaded area, we denote the initial condition in which multiple scatterings are resummed into Wilson lines and the potential is assumed to be constant.
The two orange-shaded areas represent additional rotations of the WLs and correspond to the evolution in rapidity. 
The dependence on $t$ of $\gamma_\rvec^t$ in the orange area is extracted from the evolved dipole.

Finally, let us note that in the GT, there is an exact Casimir scaling. Color factors are automatically generated by using the irrep generators $t^a_R$, {i.e.} $(t_R)^2 = C_R$. We will return to this property in section \ref{sec. scaling violation}, after we evaluate the evolution of both the fundamental dipole and adjoint dipole.

\subsection{Numerical extraction}
\label{sec. numerics}

In order to recover the function $\gamma_\rvec^t$, we will assume a reflection symmetry, $\gamma_\rvec^t = \gamma_\rvec^{-t}$ such that we have\footnote{If we consider the evolution of the fundamental dipole given by the BK equation, the {r.h.s.} is directly proportional to the {r.h.s.} of the BK equation to order $\calo(\alpha_s)$ for $\alpha_s \delta Y \ll 1$}:
\begin{equation}
\label{eq:gamma_evo_from_dipole}
    \gamma_\rvec^t = \lim\limits_{\delta Y \rightarrow 0} \ \frac{-1}{2\delta Y\, C_R} \ln \left( \frac{S_\rvec^{R,t+\delta Y}}{S_\rvec^{R,t}} \right) \geq 0
\end{equation}
where the factor $2$ follows from the reflection symmetry. Note that this expression corresponds to the continuum limit when the rapidity interval, $\delta Y$, becomes infinitesimal. For the numerical analysis, this expression will be made discrete, see \req{eq:gamma_evo_discr}.
Once this two-point function is extracted from the dipole, one can evaluate the Gaussian truncation of any operator of the Balitsky hierarchy by simply using Wick contractions.

In practice, we do not compute all our observables after every step of rapidity evolution. Instead, we extract the expectation value of those operators at fixed rapidity given by: 
\begin{equation}
    y = n \times (\alpha_s\delta Y), \qquad n\in\{0,1,2,\cdots 19\}, \qquad \alpha_s \delta Y = 1/20.
\end{equation}
Note that the WLs are evolved in the same range up to $\alpha_s Y_{\text{max}}$; however, the corresponding rapidity step $\delta_y$ is smaller 
\begin{equation}
    \alpha_s\delta_y = 0.01 
\end{equation}
such that we have $\alpha_s \delta_y \ll \alpha_s \delta Y \ll 1$.

\subsection{Fit parametrizations}
\label{sec. fits}

In order to facilitate further calculations, and in particular, to enable the analytic determination of derivatives of $\gamma_\rvec^t$, we model the data through two-parameter fits for the initial condition and three-parameter fits for each evolution slice.

The initial condition for the dipole, obtained through the numerical evaluation of the (massive) MV-model, is fitted to the following function
\begin{equation}\label{eq:fit_model_IC}
    S_r^{R,Y=0} \Big|_\text{fit} = e^{-C_R \Gamma_r^{(\text{ic fit})}}, 
    \qquad \text{where} \qquad
    \Gamma_r^{(\text{ic fit})} = \frac{\mu^2_{\text{fit}}}{m^2_{\text{fit}}}\big[1- m_{\text{fit}} r K_1(m_{\text{fit}}r) \big].
\end{equation}
The function $K_1$ is the modified Bessel function of the second kind.
The capital $\Gamma_r^{(ic)}$ is related to the previous $\gamma_r^t$ using
\begin{equation}
    \Gamma_r^{(ic)} = \int\limits_{-1}^{1} dt\ \gamma_r^t = 2 \gamma_r^0
\end{equation}
which follows from our assumption that there is no dependence on the longitudinal profile at the initial condition.

\begin{figure}
    \centering
    \includegraphics[width=0.49\linewidth]{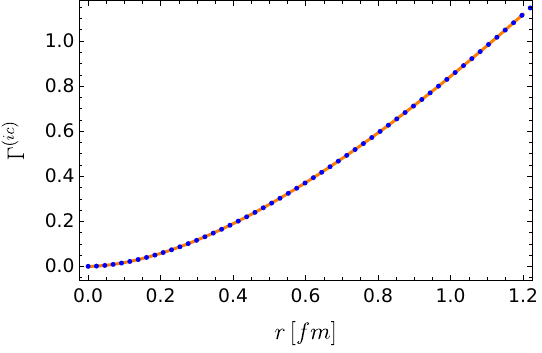}
    \hfill
    \includegraphics[width=0.50\linewidth]{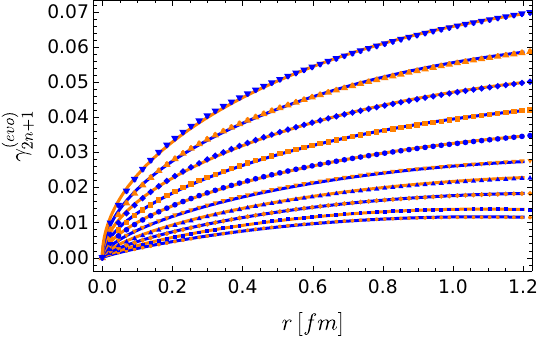}
    \caption{
    Left: the logarithm of the (fundamental) dipole rescaled by $1/C_F$ at the initial condition. Data points are generated according to the MV-model with parameters $\mu=0.7\,\text{fm}^{-1},\ m=0.2\,\text{fm}^{-1},\ L=100\,\text{fm}$ on a grid of $6144^2$ nodes; the fit, represented in plain line, is performed according to \req{eq:fit_model_IC}.\\
    Right: The change in the (rescaled by $1/C_F$ fundamental) dipole is characterized by $\gamma_r^t$ given in \req{eq:gamma_evo_from_dipole} for a fixed value of $\alpha_s \delta Y = 0.05$. Fits are performed according to \req{eq:fit_model_Evo} and are represented in plain lines. For clarity of the figure, we only show the odd extraction of $\gamma^{(evo)}(r)$ where the curves increase in magnitude along with the rapidity label.
    }
    \label{fig:Fit_fun_dipole}
\end{figure}

\begin{figure}
    \centering
    \includegraphics[width=0.49\linewidth]{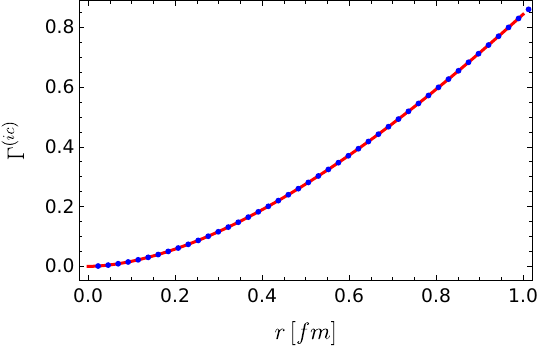}
    \hfill
    \includegraphics[width=0.50\linewidth]{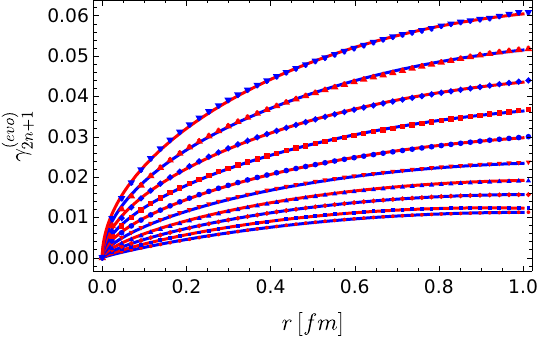}
    \caption{
    Left: the logarithm of the (adjoint) dipole rescaled by $1/C_A$ at the initial condition. Data points are generated according to the MV-model with parameters $\mu=0.7\,\text{fm}^{-1},\ m=0.2\,\text{fm}^{-1},\ L=100\,\text{fm}$ on a grid of $6144^2$ nodes; the fit, represented in plain line, is performed according to \req{eq:fit_model_IC}.\\
    Right: The change in the (rescaled by $1/C_A$ adjoint) dipole is characterized by $\gamma_r^t$ given in \req{eq:gamma_evo_from_dipole} for a fixed value of $\alpha_s \delta Y = 0.05 $. Fits are performed according to \req{eq:fit_model_Evo} and are represented in plain lines. For clarity of the figure, we only show the odd extraction of $\gamma^{(evo)}(r)$ where the curves increase in magnitude along with the rapidity label.
    }
    \label{fig:Fit_adj_dipole}
\end{figure}

The variation of the dipole between two rapidity slices $Y_{n+1} - Y_n = \delta Y$, which we denote by $S_r^{R,n}$ and $S_r^{R,n+1}$, is extracted using
\begin{equation}\label{eq:gamma_evo_discr}
    \gamma_r^{(evo),n} = \frac{-1}{2 \delta Y\ C_R} \ln \left( \frac{S_r^{R,n+1}}{S_r^{R,n}}\right),
\end{equation}
which we fit to the following ansatz:
\begin{equation}\label{eq:fit_model_Evo}
    \gamma^{(evo),n}_r\Big|_{\text{fit}} = b_n r^2 + c_n r^{\delta_n}.
\end{equation}
Note that we enforce in our ansatz choice $\delta_n > 0$ such that we automatically have
\begin{equation}
    \forall n, \quad \lim\limits_{r\rightarrow 0} \gamma^{(evo),n}_r = 0 
\end{equation}
We show in {fig.~\ref{fig:Fit_fun_dipole}} (resp. {fig.~\ref{fig:Fit_adj_dipole}}) the extraction of both $\Gamma_r^{(ic)}$ and $\gamma_r^{(evo),n}$, for all odd $n$, to be used in the quark-gluon (resp. gluon-gluon) sector for the evaluation of the TMD distributions. 
The TMD distribution operator definition can be found in the appendix \ref{App:operators}.
{The fit parameters can be found in the appendix \ref{App:fit}.}

The value of $R_\text{cut} = 1.2\, \text{fm}$ (resp. $1.0\,\text{fm}$) is chosen {w.r.t.} the statistical errors of our simulations. 
For a fixed separation $r > R_\text{cut}$, the mean value of the data for $\gamma_r^{(evo),y}$ becomes non-monotonous {w.r.t.} the change in rapidity $y$, thus we reject this region for our fit.
We expect an increase in the number of realizations to allow for a larger range than $1.2\, \text{fm}$ (resp. $1.0\,\text{fm}$); however, the scope of this paper does not require such accuracy.
Our focus is on features appearing around the saturation radius at the initial condition, and those features drift toward the ultraviolet as we evolve in rapidity.

Finally, let us write the dipole at the rapidity step $ Y = n $ as
\begin{equation}
    S_r^{R,Y=n} = \exp \left[ - C_R \left( \Gamma_r^{(ic)} + \sum_{\ell=1}^n 2\,\gamma^{(evo),\ell}_r \right) \right],
    \label{eq. reconstructed dipole}
\end{equation}
which can be used as a cross-check of the extraction method. We do recover perfect agreement between $S_r^{R,Y=n}$ given by \ref{eq. reconstructed dipole} and both the fundamental and adjoint dipole obtained from direct solutions of the JIMWLK equation in the full range of rapidities and separations $r$ considered.

\subsection{Casimir scaling under evolution}
\label{sec. scaling violation}

\begin{figure}
    \centering
    \includegraphics[width=0.95\linewidth]{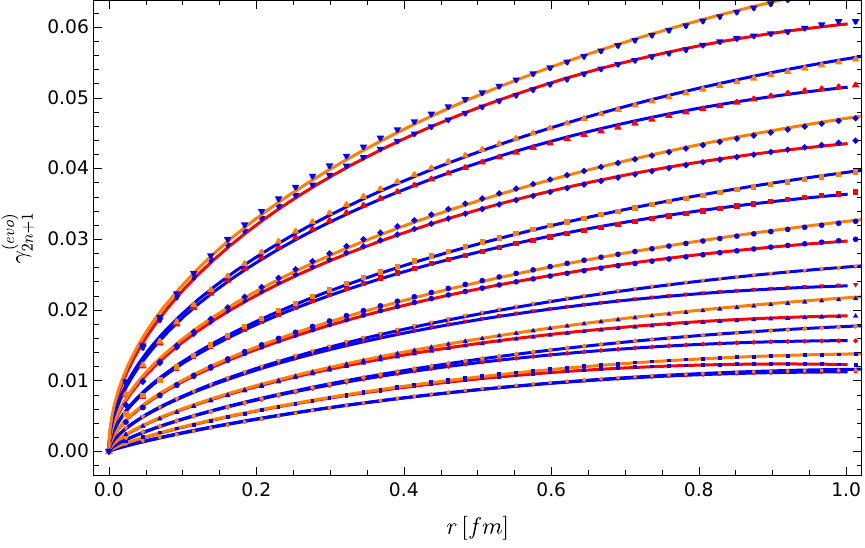}
    \caption{
    We superpose the right panel of {fig.~\ref{fig:Fit_fun_dipole}} and {fig.~\ref{fig:Fit_adj_dipole}}.
    The slight deviation of the data points (and curves) between the fundamental dipole and adjoint dipole hints at a mild breaking of Casimir scaling under evolution in the region of interest.
    }
    \label{fig:Casimir_scaling}
\end{figure}

From the extraction of $\gamma_r^{(evo)}$ for both the fundamental and adjoint dipole, one can check the validity of the Casimir scaling.
In {fig.~\ref{fig:Casimir_scaling}}, we show the extraction of $\gamma_r^{(evo),n}$ for every odd number $n$. Colors are alternating between blue and orange (resp. blue and red) when the fundamental (resp. adjoint) dipole is used.

At the beginning of the evolution (bottom curves), we see a good agreement between both extractions and thus support Casimir scaling. 
Toward the end of the evolution (top curves), we see a growing gap between the two extractions; see e.g., the top two curves in orange and red, which correspond to the very last extraction of $\gamma^{(evo),n}$ at $n=19$. 
From this observation, we conclude that there is a mild breaking of Casimir scaling induced by the evolution.
{It has been argued, see {e.g.} \cite{Kovchegov:2008mk}, that the size of the  Casimir scaling violations can be used to estimate the impact of multi-reggeon exchanges.}

To minimize this effect, we will use the extraction from the fundamental (res. adjoint) dipole to evaluate the GT of TMD distributions in the quark-gluon sector (resp. gluon-gluon sector) in the following sections.

\section{Results for the \texorpdfstring{$\Omega$}{Omega} operators}
\label{sec:Omega}

Having access to the two-point correlation function ${\gamma}^t_{\uvec - \vvec}$ discussed in the previous section, we can easily obtain the Gaussian truncation for the \texorpdfstring{$\Omega$}{Omega} operators that are the building blocks of the small-$x$ TMD distributions. After reminding the definitions of the \texorpdfstring{$\Omega$}{Omega} operators in section \ref{sec. omega definitions}, first, we provide in section \ref{sec. omega analytic results} the complete analytic formulae, and then, in section \ref{sec. omega numerical results}, we present the resulting curves using the fitted ${\gamma}^t_{\uvec - \vvec}$.

\subsection{Definitions}
\label{sec. omega definitions}

We study the expectation value of the operators $\widehat{\Omega}^\omega_{ag}$ (where $a$ is either $q$ for quark or $g$ for gluon) defined in \cite{Cougoulic:2026drr}, before studying the TMD distributions given in Appendix \ref{App:operators} (the latter are linear combinations of the former).
Using a birdtrack notations \cite{Cvitanovic:1976am,Cvitanovic:2008zz,Dokshitzer:1995fv,Keppeler:2017kwt,Peigne:2024srm}, these operators read \cite{Cougoulic:2026drr}:
\begin{align}
\widehat{\Omega}_{qg}^\omega &= 
{2} \times \ 
\vcenter{\hbox{
\begin{tikzpicture}[scale=0.3]
    \draw[thick] (-5,1.5) -- ++ (10,0);\draw[thick,->] (-5,1.5) -- ++ (3,0);
    \draw[thick] (-5,2) -- ++ (10,0);\draw[thick,-<] (-5,2) -- ++ (3,0);
    \draw[fill=Blue!50!white] (1,1.5) circle (0.2);
    \draw[thick] (-5,1) -- ++ (10,0);\draw[thick,-<] (-5,1) -- ++ (4,0);
    \draw[thick] (-5,-1) -- ++ (10,0);\draw[thick,->] (-5,-1) -- ++ (6,0);
    \draw[thick] (-5,-1.5) -- ++ (10,0);\draw[thick,-<] (-5,-1.5) -- ++ (7,0);
    \draw[thick] (-5,-2) -- ++ (10,0);\draw[thick,->] (-5,-2) -- ++ (7,0);
    \draw[fill=Blue!50!white] (-1,-2) circle (0.2);
    \draw[thick] (-5,2) to[out=180,in=180] (-5,1.5);
    \draw[thick] (-5,-2) to[out=180,in=180] (-5,-1.5);
    \draw[thick] (-5,1) to[out=180,in=180] (-5,-1);
    \draw[thick] (5.5,2) to[out=0,in=90,looseness=1.5] (6,1.75) to [out=-90,in=0, looseness=1.5] (5.5,1.5);
    \draw[thick] (5.5,-2) to[out=0,in=-90,looseness=1.5] (6,-1.75) to [out=90,in=0, looseness=1.5] (5.5,-1.5);
    \draw[gluon] (6,1.75) to[out=0,in=0,looseness=2] (6,-1.75);
    \draw[thick] (5.5,1) -- (6,1) to[out=0,in=0,looseness=2] (6,-1) -- (5.5,-1);
    \draw[fill=white] (7.5,0) ellipse (1.2 and .5) node {\scriptsize $\omega$};
\end{tikzpicture}}}, \\
\widehat{\Omega}_{gg}^\omega &= {4} \times
\vcenter{\hbox{
\begin{tikzpicture}[scale=0.3]
    \draw[thick] (-5,2) -- ++ (10,0);\draw[thick,->] (-5,2) -- ++ (3,0);
    \draw[thick] (-5,2.5) -- ++ (10,0);\draw[thick,-<] (-5,2.5) -- ++ (3,0);
    \draw[fill=Blue!50!white] (1,2) circle (0.2);
    \draw[thick] (-5,1.5) -- ++ (10,0);\draw[thick,-<] (-5,1.5) -- ++ (4,0);
    \draw[thick] (-5,1) -- ++(10,0);\draw[thick,->] (-5,1) -- ++(4,0);
    \draw[thick] (-5,-1) -- ++(10,0);\draw[thick,-<] (-5,-1) -- ++(6,0);
    \draw[thick] (-5,-1.5) -- ++ (10,0);\draw[thick,->] (-5,-1.5) -- ++ (6,0);
    \draw[thick] (-5,-2) -- ++ (10,0);\draw[thick,-<] (-5,-2) -- ++ (7,0);
    \draw[thick] (-5,-2.5) -- ++ (10,0);\draw[thick,->] (-5,-2.5) -- ++ (7,0);
    \draw[fill=Blue!50!white] (-1,-2.5) circle (0.2);
    \draw[thick] (-5,2) to[out=180,in=180] (-5,2.5);
    \draw[thick] (-5,-2) to[out=180,in=180] (-5,-2.5);
    \draw[thick] (-5,1) to[out=180,in=180] (-5,-1);
    \draw[thick] (-5,1.5) to[out=180,in=180] (-5,-1.5);
    \draw[thick] (5.5,2.5) to[out=0,in=90] ++(.5,-.25) to[out=-90,in=0] ++(-.5,-.25);
    \draw[thick] (5.5,1.5) to[out=0,in=90] ++(.5,-.25) to[out=-90,in=0] ++(-.5,-.25);
    \draw[thick] (5.5,-1) to[out=0,in=90] ++(.5,-.25) to[out=-90,in=0] ++(-.5,-.25);
    \draw[thick] (5.5,-2) to[out=0,in=90] ++(.5,-.25) to[out=-90,in=0] ++(-.5,-.25);
    \draw[gluon] (6,2.25) to[out=0,in=0,looseness=2] (6,-2.25);
    \draw[gluon] (6,1.25) to[out=0,in=0,looseness=2] (6,-1.25);
    \draw[fill=white] (8,0) ellipse (1.2 and .5) node {\scriptsize $\omega$};
\end{tikzpicture}}}.
\end{align}
Horizontal lines are WLs at the position $\xvec$ (top) or $\yvec$ (bottom), and vertical lines denote contractions at $t = -\infty$ (left) or $t = + \infty$ (right). The blob on the right denotes the projection onto the irrep $\omega$ which belongs to either the decomposition $\bar{3} \otimes 8$ ($a=q$) or $8 \otimes 8$ ($a=g$). Pre-factors follow from the normalization convention for the fundamental generators, which is $\tr (t^a t^b) = \thalf \delta^{ab}$.
The blue bullets on the WLs denote the action of $\nabla^j_\xvec$ or $\nabla^j_\yvec$ on the corresponding WL.

\subsection{Analytic results for the Gaussian truncation}
\label{sec. omega analytic results}

The expectation values of the $\widehat{\Omega}^\omega_{ag}$ operators receive two contributions, denoted $(a)$ and $(b)$. 

\paragraph{The contribution $(b)$.} 
It involves the diagrams with the correlation $\langle (\partial^j\alpha)_\xvec (\partial^j \alpha)_\yvec \rangle $ from Wick contractions, see \cite{Cougoulic:2026drr} for more details.
It is the only contribution for all irrep $\omega$ such that
\begin{subequations}
\begin{align}
    \forall \omega \in \bar{3} \otimes 8,\, \omega \neq \bar{3}\ :
    \qquad
    &\Omega_{qg}^\omega = \Omega_{ag,\omega}^{(b)}, \\
     \forall \omega \in 8 \otimes 8,\, \omega \neq 8_a:
    \qquad
    &\Omega_{gg}^\omega = \Omega_{gg,\omega}^{(b)}.
\end{align}
\end{subequations}
We can write in the Gaussian truncation the following expectation value \cite{Cougoulic:2026drr}:
\begin{align}\label{eq:omega_ag_B}
\Omega_{ag,\omega}^{(b)}
    &= - \frac{K_\omega}{2} \int\limits_{-\overline{Y}}^{\overline{Y}} dt\ e^{-C_\omega \int^{\overline{Y}}_t dc\ \gamma_r^c} \left( \nabla^2_r \gamma_r^t\right) e^{-C_R \int^t_{-\overline{Y}} db\ \gamma_{r}^{b} },
\end{align}
where $a=q$ implies $C_R = C_F$ and $a=g$ implies $C_R = C_A$. 
$K_\omega$ is the dimension of the irrep $\omega$ in the decomposition $R \otimes 8$.

\noindent
Using reflection symmetry $\gamma_r^t = \gamma_r^{-t}$ and the invariance over the longitudinal profile of $\gamma^t_\rvec$ in the initial condition, one finds after some algebra
\begin{align}\label{eq:omega_ag_B_after_algebra}
\Omega_{ag,\omega}^{(b)}
    = &- \frac{K_\omega}{2}  \exp\left\{-\frac{C_\omega + C_R}{2} \int^{\overline{Y}}_{0} da\ 2\gamma_{r}^a \right\} \times \Big[
    \left( \nabla^2_r\, \Gamma_r^{(ic)}\right) \text{sinhc}\left(\tfrac{C_R-C_\omega}{2}\Gamma_r^{(ic)} \right) \notag \\
    &+
    \int\limits_{1}^{\overline{Y}} dt\
    \left( \nabla^2_r \gamma_{r}^t\right) 
    \times  \left(e^{-\frac{C_R-C_\omega}{2} \int_{0}^{|t|}dc\, 2\gamma_r^c} + e^{-\frac{C_\omega-C_R}{2} \int_{0}^{|t|}dc\, 2\gamma_r^c}  \right)
    \Big].
\end{align}
In this expression, we recognize some mixing between operators. On the one hand, we have the TMD operator expectation value evaluated at the initial condition given by
\begin{equation}
    \frac{-K_\omega}{2}e^{-\frac{C_\omega+C_R}{2}\Gamma_r^{(ic)}}\left( \nabla^2_r\, \Gamma_r^{(ic)}\right) \text{sinhc}\left(\tfrac{C_R-C_\omega}{2}\Gamma_r^{(ic)} \right).
\end{equation}
This contribution is then evolved according to regular eikonal evolution denoted by the overall prefactor
\begin{equation}
    \exp\left\{-\frac{C_\omega + C_R}{2} \int^{\overline{Y}}_{1} da\ 2\gamma_{r}^a \right\},
\end{equation}
up to rapidity $Y= \overline{Y}-1$.
On the other hand, we have a mixing where we start with a dipole operator in either the irrep $R$ or $\omega$, which is evolved up to the rapidity $t - \delta_y$.    
At the rapidity $t$, it mixes into the TMD operator and is then evolved according to the eikonal evolution up to rapidity $Y = \overline{Y}-1$.

Importantly, in the expression \req{eq:omega_ag_B_after_algebra}, one still needs the information of the whole dipole evolution. This situation is distinct from the cases explored in \cite{Lappi:2026kxi}, where the dependence on the rapidity parametrization $y$ disappeared in favor of $\Gamma_{\uvec\,\vvec}$ being the potential integrated over the longitudinal profile.

\paragraph{The contribution $(a)$.} 
There are no irrep $\omega$ for which $\Omega^{(a)}_{ag,\omega}$ appears without $\Omega^{(b)}_{ag,\omega}$. The sum of the two contributions $\Omega^{(a+b)}_{ag}$ is simply given by
\begin{equation}\label{eq:omega_ab}
    \Omega^{(a+b)}_{qg,3} = \frac{K_F}{2C_F} \Delta_r S_r, 
    \qquad 
    \Omega^{(a+b)}_{gg,8a} = \frac{K_A}{2C_A} \Delta_r S^{(adj)}_r.
\end{equation}
While the content in terms of correlators resulting from Wick contractions appears to be richer: one has to consider both $\langle \alpha(\partial\alpha)\rangle$ and $\langle (\partial^j\alpha) (\partial^j \alpha) \rangle $ in addition to the common $\langle \alpha \alpha\rangle$, they combine into the action of the Laplace operator on the dipole. {We anticipate that the two TMD distributions related to the Laplacian of the dipoles will be accurately reproduced by the Gaussian truncation.}

\subsection{Numerical results}
\label{sec. omega numerical results}

We are now in a position to compare the predictions of the Gaussian truncation with the data generated from the JIMWLK evolution equation. We start by discussing the operators from the quark-gluon sector, and afterwards turn to the operators from the gluon-gluon sector.

\subsubsection{Quark-Gluon sector}

We have three expectation values to consider in the quark-gluon sector, $\Omega_{qg, \omega}$, $\omega \in \{3, 6, 15\}$. We start with  $\Omega_{qg, 3}$, which is directly related to the fundamental dipole distribution by the Laplacian operator.

\begin{figure}
    \centering
    \includegraphics[width=0.95\linewidth]{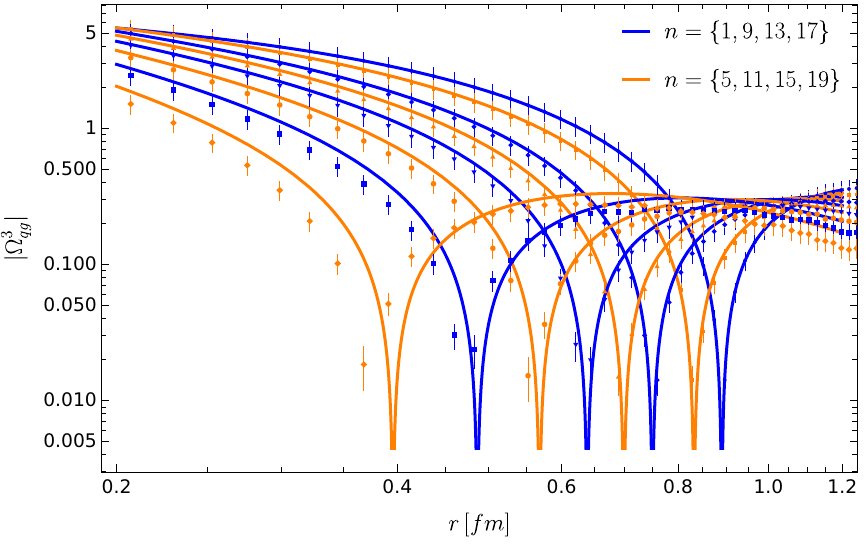}
    \caption{
    The expectation value $|\langle \Omega_{qg}^3 \rangle |$ for odd rapidity steps around the location of the node of the distribution. 
    As the rapidity $n\, \alpha_s\delta Y$ increases, the distribution drifts toward the UV (colors are alternated between blue and orange according to the figure labels). 
    \PointsLines
    }
    \label{fig:Omega_qg_3_node-region}
\end{figure}

Figure~\ref{fig:Omega_qg_3_node-region} shows the numerical data obtained from the simulation for the absolute value of the expectation value of the operator $\widehat{\Omega}_{qg}^3$ (data points), alongside the result of the Gaussian truncation (plain lines). 
We observe that the GT is in good agreement with the numerical data.
As discussed in the previous section, the dipole is the input to the Gaussian truncation, and since ${\Omega}_{qg}^{3}$ is directly related to the Laplacian of the fundamental dipole (see \req{eq:omega_ab}), it is not surprising that the GT provides a good estimate of the numerical data.
The good agreement observed is thus simply a consequence of the quality of the extraction of $\gamma_\rvec^t$, as shown in {Fig.~\ref{fig:Fit_fun_dipole}}.

The agreement in {Fig.~\ref{fig:Omega_qg_3_node-region}} is not perfect. 
There are clearly some deviations towards the end of the evolution, where the GT results are slightly slower than the data. See, for example, the data points and curves for $n=\{15,17,19\}$.
This issue can be resolved by extracting $\gamma_\rvec^t$ from both the dipole and  ${\Omega}_{qg}^{3}$.\footnote{{
Since we evaluate operators using JIMWLK evolution, we naturally have access to the expectation values of both the dipole and ${\Omega}_{qg}^{3}$. Thus, our framework enables us to simultaneously fit both operators' expectation values.
However, in order to take advantage of the GT and completely bypass the computationally intensive JIMWLK simulations, one needs a closed equation for both the dipole and ${\Omega}_{qg}^{3}$, with the BK equation being the obvious candidate. It should be kept in mind that an accurate extraction of the Laplacian of the dipole is required, and thus methods such as automatic differentiation \cite{Cougoulic:2024jnd} should be preferred.
}}
This would allow for a better agreement in the location of the node without deteriorating the quality of the results shown in {Fig.~\ref{fig:Fit_fun_dipole}}.
This approach was not chosen here, since the consistency of approach was a concern and the relation between the dipole and ${\Omega}_{qg}^{3}$ under the GT was to be checked by predicting ${\Omega}_{qg}^{3}$ from the dipole data.
However, a simultaneous fit should be preferred in a quantitative analysis.
One could track the location of the node of this distribution to indicate the evolution speed of this operator, which is essential for making systematic comparisons of the evolution of different operators.
We conclude that the GT is a robust approximation for $\Omega_{qg}^3$.

\begin{figure}
    \centering
    \includegraphics[width=0.49\linewidth]{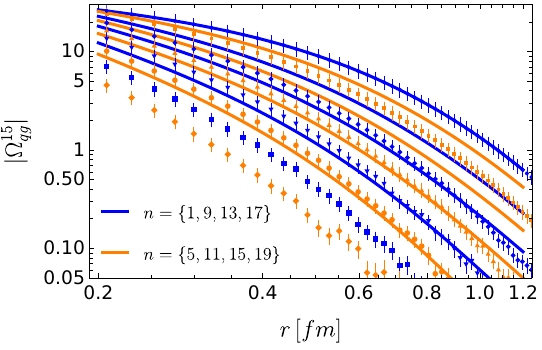}
    \hfill
    \includegraphics[width=0.49\linewidth]{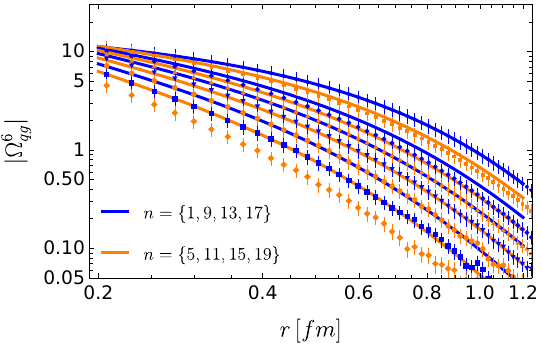}
    \caption{
    The expectation value $|\langle \Omega_{qg}^{15} \rangle |$ (left) and $|\langle \Omega_{qg}^{6} \rangle |$ (right) in the range $r \in (0.2\,\text{fm}, 1.2\,\text{fm})$ common to {fig.~\ref{fig:Omega_qg_3_node-region}}. 
    As the rapidity $n\, \alpha_s\delta Y$ increases, distributions drift toward the UV (colors are alternated between blue and orange according to the figure labels).
    \PointsLines
    }
    \label{fig:Omega_15_6_node-range}
\end{figure}

Figure \ref{fig:Omega_15_6_node-range} shows the numerical data obtained from the simulation for the operators $\widehat{\Omega}_{qg}^{15}$ and $\widehat{\Omega}_{qg}^{6}$ (data points), alongside the corresponding Gaussian truncation (plain lines). 
The functional form of both distributions appears to be well captured using the GT.
However, the data clearly favor a faster evolution than can be described by the GT.
This is evident in the final rapidity labeled $n=19$. 
In the left panel, the GT result for $\widehat{\Omega}_{qg}^{15}$ at $n=19$ is closer to the data at $n=15$.
In the right panel, the GT result for $\widehat{\Omega}_{qg}^{6}$ at $n=19$ is closer to the data at $n=17$.

Interestingly, there appears to be an ordering of the mismatch between data and GT according to the representation involved: $\omega = \{3,6,15\}$.
The largest discrepancy is for the largest irrep, $\omega = 15$, followed by the irrep $\omega = 6$, while the fundamental irrep shows a relatively good agreement.
This observation hints at a potential solution to this mismatch by effectively rescaling the rapidity according to the charge or dimension of the irrep involved: $Y \rightarrow Y_\omega$, where $Y_\omega$ would be the rescaled rapidity to be used for the expectation value of the operator $\widehat{\Omega}_{qg}^\omega$ in the Gaussian truncation.


\subsubsection{Gluon-Gluon sector}
\label{subsec:omega_gg}

\begin{figure}
    \centering
    \includegraphics[width=0.95\linewidth]{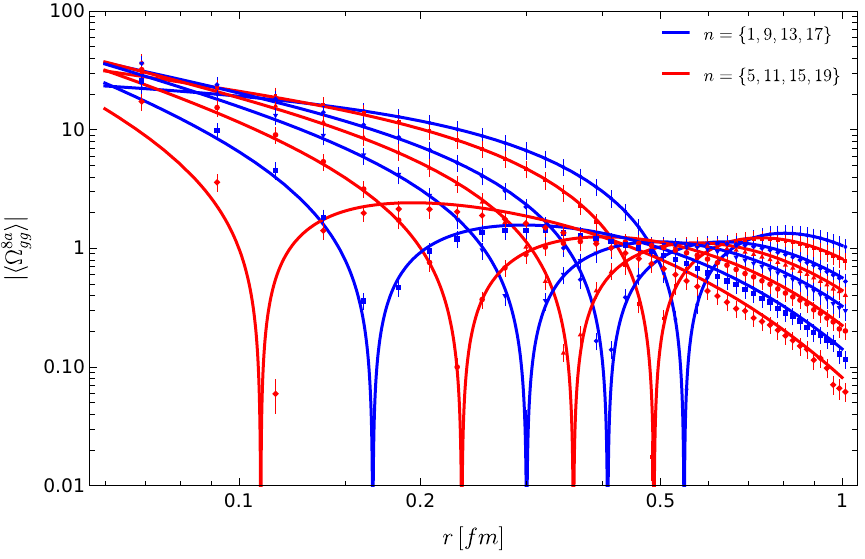}
    \caption{
    The expectation value $|\langle \Omega_{gg}^{8a} \rangle |$ around the location of the node of the distribution. 
    As the rapidity $n\, \alpha_s\delta Y$ increases, the distribution drifts toward the UV (colors are alternated between blue and red according to the figure labels). 
    \PointsLinesAdj
    }
    \label{fig:Omega_gg_8a_node-region}
\end{figure}

\begin{figure}
    \centering
    \includegraphics[width=0.49\linewidth]{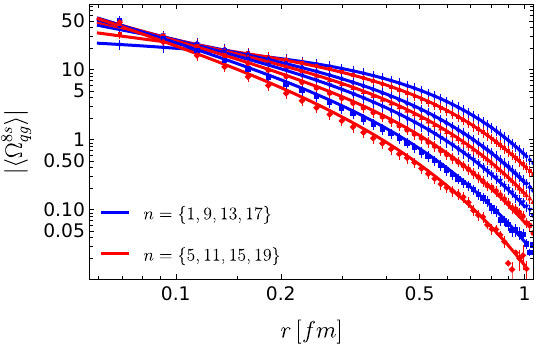}
    \hfill
    \includegraphics[width=0.49\linewidth]{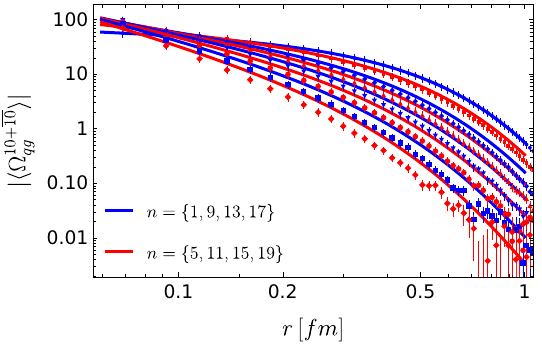}
    \vfill
    \includegraphics[width=0.49\linewidth]{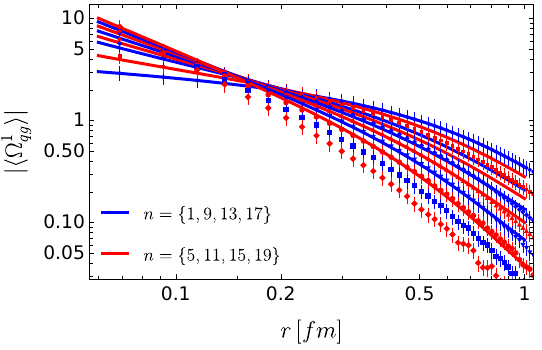}
    \hfill
    \includegraphics[width=0.49\linewidth]{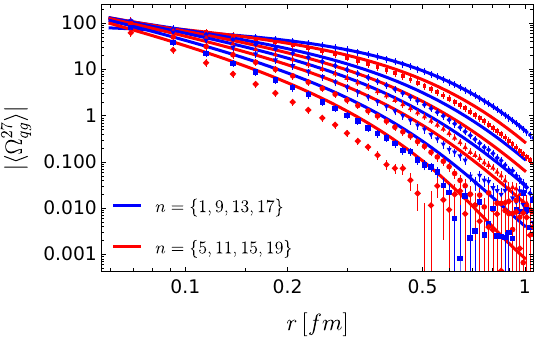}
    \caption{
    Top: The expectation value $|\langle \Omega_{gg}^{8s} \rangle |$ and $|\langle \Omega_{gg}^{10+\overline{10}} \rangle |$.
    Bottom: The expectation value $|\langle \Omega_{gg}^{1} \rangle |$ and $|\langle \Omega_{gg}^{27} \rangle |$.
    As the rapidity $n\, \alpha_s\delta Y$ increases, distributions drift toward the UV (colors are alternated between blue and red according to the figure labels). 
    \PointsLinesAdj
    }
    \label{fig:Omega_8s_20_27_1_node-range}
\end{figure}

We would like to emphasize that in the gluon-gluon sector, we use the adjoint dipole to determine the function $\gamma_\rvec^t$.
The extraction is shown in {fig.~\ref{fig:Fit_adj_dipole}} and $\gamma_\rvec^t$ is used in the GT curves shown in {fig.~\ref{fig:Omega_gg_8a_node-region}} and in {fig.~\ref{fig:Omega_8s_20_27_1_node-range}}.
Figure~\ref{fig:Omega_gg_8a_node-region} shows the numerical data obtained from the simulation for the absolute value of the expectation value of the operator $\widehat{\Omega}_{gg}^{8_a}$ (data points), alongside the result of the Gaussian truncation (plain lines). 
As with $\widehat{\Omega}_{qg}^3$ in the quark-gluon sector, the expectation value of $\widehat{\Omega}_{gg}^{8_a}$ is related to the Laplacian of the adjoint dipole and is well described by the Gaussian approximation.
The location of the node, which can be used to estimate the speed of evolution of this operator, is also well described by the GT.

The top left panel of {fig.~\ref{fig:Omega_8s_20_27_1_node-range}} shows the expectation value of the operator $\widehat{\Omega}_{gg}^{8_s}$.
There is a good agreement between the simulation data and the GT results across the rapidity range.
While the irrep $8_a$ depends on $\Omega_{gg}^{(a+b)}$ given by \req{eq:omega_ab}, the irrep $8_s$ depends only on $\Omega_{gg}^{(b)}$ given by \req{eq:omega_ag_B_after_algebra}.
It is worth noting that, for $\omega = 8_s$, $C_R = C_A = C_\omega$. 
Therefore \req{eq:omega_ag_B_after_algebra} simplifies greatly and the result only depends on the integrated potential:
\begin{equation}
    \Gamma^Y_r = \int\limits_{-(1+Y)}^{1+Y} dt\ \gamma_r^t.
\end{equation}
The entire rapidity dependence of the dipole is lost in favor of its value at the rapidity $Y$, a shared property with $\Omega_{gg}^{8_a}$.

In the right-hand column of {fig.~\ref{fig:Omega_8s_20_27_1_node-range}}, we show the expectation values of the operator $\widehat{\Omega}_{gg}^{10+\overline{10}}$ and of the operator $\widehat{\Omega}_{gg}^{27}$. 
Those features are reminiscent of the previously discussed irreps $\omega = \{15,6\}$ in the quark-gluon sector.
While the functional form appears to be correct, the evolution given by the GT is slower than the data.
The data points for $n = 17$ can be compared with the GT curves for $n=19$.  
As in the quark-gluon case, an ordering can be observed: the discrepancy between the data points and the GT curves appears to be greater with the irrep $\omega = 27$ than with the irrep $\omega = 10+\overline{10}$, compared with the irreps $\omega = 8_{a}$ and $\omega = 8_{s}$, which agree with the data.

Finally, the bottom left panel of {fig.~\ref{fig:Omega_8s_20_27_1_node-range}} shows the expectation value of the operator $\widehat{\Omega}_{gg}^1$. 
As in all previous cases, we also observe that, for $\omega=1$, the rapidity dependence of the GT curves is slower than the data. For example, the GT curve at $n=19$ appears to agree with the data at $n=15$.
This pattern is observed in the quark-gluon sector and in the gluon-gluon sector for irreps $\omega = \{27, 10+\overline{10},8_s,8_a \}$, where the discrepancy between the simulation data and the GT curves increases with the irreps involved, with agreement observed for the adjoint (resp. fundamental) in the gluon-gluon (resp. quark-gluon) sector.
Since $K_1 < K_A$ and $C_1 < C_A$, one might have expected the GT curves to be faster than the data points, not slower!
This puzzling observation discredits a naive rescaling of the rapidity $Y \rightarrow Y_\omega$ to effectively improve the agreement between simulation data and the GT. 
We cannot explain this last observation, and further studies are required to understand this effect.

\subsection{Summary}
We find a good agreement between the JIMWLK-evolved operators $\widehat{\Omega}_{ag}^\omega$ and their GT prediction, for the irrep $\omega = 3$ in the quark-gluon sector and $\omega = \{8_a, 8_s\}$ in the gluon-gluon sector.
While we expected a robust agreement for $\omega = 3$ (resp. $\omega = 8_a$) due to the operator being related to the Laplacian of the fundamental (resp. adjoint) dipole, we also find a very good agreement for the expectation value of the operator $\widehat{\Omega}_{gg}^{8_s}$, which depends on the contribution $\Omega_{gg}^{(b)}$. In this case, the color charges involved in the expression of $\Omega_{gg}^{(b)}$ conspire such that the result depends only on the integrated potential $\Gamma_r^Y$ as opposed to the full rapidity dependence of $\gamma_r^y$.
For all other irreps involved in the decompositions of $3 \otimes 8$ and $8 \otimes 8$ that were not previously mentioned, we observe that the GT underestimates the evolution speed of the JIMWLK data.
The discrepancy between the GT prediction and the data seems to be ordered according to the dimension or the charge of the irrep $\omega$ involved.
The singlet case in the gluon-gluon sector is the only one that does not fulfill this ordering.
Like all other irreps, the contribution $\Omega_{gg}^1$ shows that the data is faster than the GT prediction.
Understanding how this discrepancy scales in terms of irreps may allow for an effective prescription to improve GT accuracy by rescaling rapidity.


\section{Results for the TMD distributions}
\label{sec:tmd}

Although we have already discussed the GT for all the $\Omega_{ag}^{\omega}$ functions, which are the building blocks of the TMD distributions given in appendix \ref{App:operators}, demonstrating that the GT is not reproducing the numerical data correctly, the conclusions for the TMD distributions are even more striking. This is because the TMD distributions are linear combinations of $\Omega_{ag}^{\omega}$ and cancelations occur, producing various structures completely absent in the data.

In order to relate the TMD distribution $\calf_{ag}^{(i)}$ to the expectation value of $\Omega_{ag}^\omega$ we use the results from \cite{Cougoulic:2026drr} for $SU(3)$. In the quark-gluon sector, it reads\footnote{Note that there are some typo in the matrix given by 5.3 of \cite{Cougoulic:2026drr}. The expressions given in 5.2a and 5.2b are correct, and we write the corrected matrix for $SU(N_c=3)$.}
\begin{equation}
\begin{pmatrix}
    \calf_{qg}^{(1)} \\
    \calf_{qg}^{(2)} \\
    \calf_{qg}^{(3)} \\
\end{pmatrix}
=
\begin{pmatrix}
    - \tfrac{8}{3} & 0 & 0\\
    - \tfrac{1}{3} & - \tfrac{1}{3} & - \tfrac{1}{3} \\
    \tfrac{1}{3} & 1 & - 1
\end{pmatrix}
\cdot
\begin{pmatrix}
    \Omega_{qg}^3 \\
    \Omega_{qg}^6 \\
    \Omega_{qg}^{15}
\end{pmatrix},
\end{equation}
while in the gluon-gluon sector we have
\begin{equation}
\begin{pmatrix}
    \calf_{gg}^{(1)} \\
    \calf_{gg}^{(2)} \\
    \calf_{gg}^{(3)} \\
    \calf_{gg}^{(4)} \\
    \calf_{gg}^{(5)} \\
    \calf_{gg}^{(6)} \\
    \calf_{gg}^{(7)}
\end{pmatrix}
=
\begin{pmatrix}
 -\frac{8}{9} & -\frac{1}{2} & -\frac{5}{18} & 0 & 0  & -\frac{1}{9} \\
 -\frac{8}{9} & \frac{1}{2} & -\frac{5}{18} & 0  & 0 & -\frac{1}{9} \\
 0 & 0 & 0 & 0 & 0  & -1 \\
 -8 & 0 & 0 & 0 &  0 & 0 \\
 -1 & 1 & -1 & 1 & -1  & 0 \\
 -\frac{1}{9} & -\frac{1}{9} & -\frac{1}{9} & -\frac{1}{9} & -\frac{1}{9} & -\frac{1}{9} \\
 \frac{1}{9} & 0 & \frac{2}{9} & 0 & -\frac{1}{3} & -\frac{1}{9}
\end{pmatrix}
\cdot
\begin{pmatrix}
    \Omega_{gg}^1 \\[.3em]
    \Omega_{gg}^{8_a} \\[.3em]
    \Omega_{gg}^{8_s} \\[.3em]
    \Omega_{gg}^{10+\overline{10}} \\[.3em]
    \Omega_{gg}^{27} \\[.3em]
    8\, \Omega_{gg}^{1} 
\end{pmatrix}.
\end{equation}

\subsection{Quark-Gluon sector}

The TMD distribution $\calf_{qg}^{(1)}$ is not shown since is it simply proportional to $\Omega_{qg}^3$; the corresponding simulation data and GT curves are given in {fig.~\ref{fig:Omega_qg_3_node-region}}

\begin{figure}
    \centering
    \includegraphics[width=0.9\linewidth]{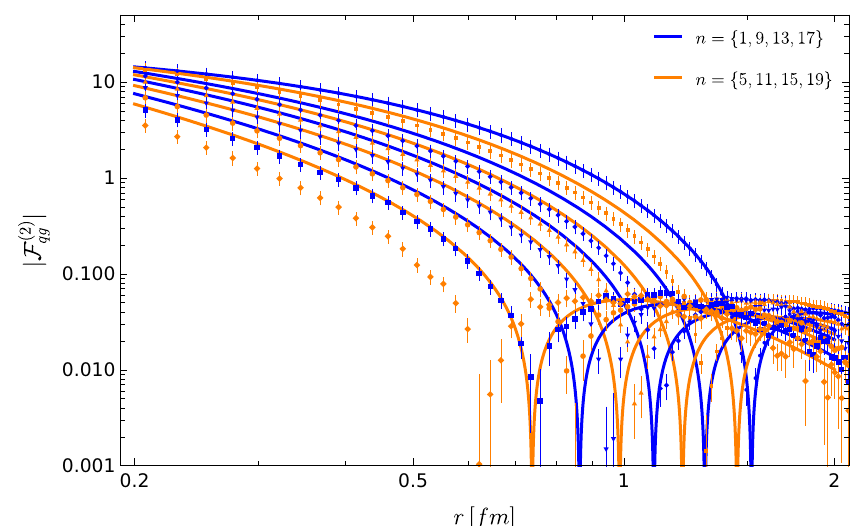}
    \hfill
    \includegraphics[width=0.9\linewidth]{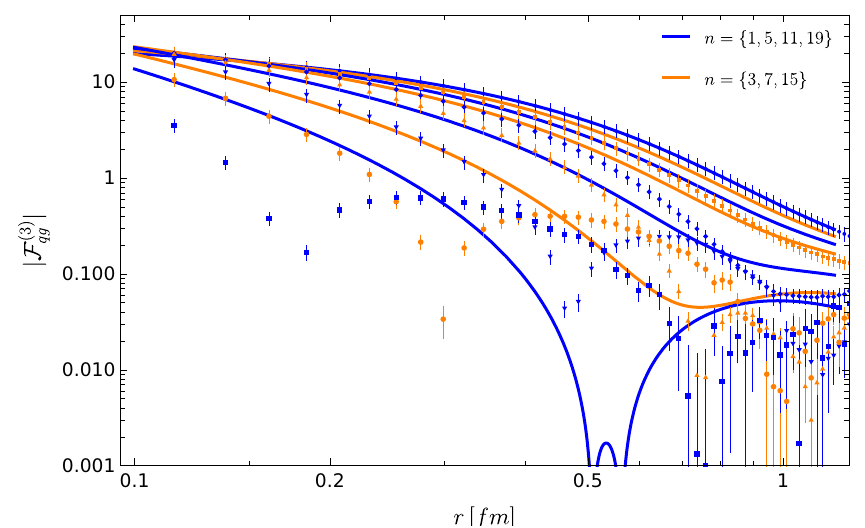}
    \caption{
    The TMD distributions $\calf_{qg}^{(2)}$ (top) and $\calf_{qg}^{(3)}$ (bottom).
    As the rapidity $n\, \alpha_s\delta Y$ increases, the distribution drifts toward the UV (colors are alternated between blue and orange according to the figure labels).
    The top panel shows $|\calf_{qg}^{(2)}|$.
    The bottom panel shows $|\calf_{qg}^{(3)}|$,
    \PointsLines
    }
    \label{fig:GT_comp_QG}
\end{figure}

The top panel of {fig.~\ref{fig:GT_comp_QG}} illustrates the TMD distributions $\calf_{qg}^{(2)}$.
While the GT curves' functional form is satisfactory, their rapidity evolution is slower than the simulation data, as evidenced by the node's location.
This behavior is consistent with the results obtained in the previous section, which showed that the rapidity evolution of both $\Omega_{qg}^6$ and $\Omega_{qg}^{15}$ given by the GT is slower than the simulation data.

Even more interesting is the discrepancy between the simulation data and the GT for $\calf_{qg}^{(3)}$ given in the bottom panel of {fig.~\ref{fig:GT_comp_QG}}.
The simulation data shows that, during rapidity evolution, the TMD $\calf_{qg}^{(3)}$ acquires a node between the rapidity labels $n=5$ (second set of blue data points) and $n=7$ (second set of orange data points). This node is preserved throughout the rapidity evolution.
Note that we plot the absolute value of the TMD $\calf_{qg}^{(3)}$ on a log-log scale; thus, the node appears as an abrupt dip between two branches. The first branch, at small $r$, has positive values while the second branch, at large $r$, has negative values.
The GT curves exhibit a different behavior as they evolve. 
Initially, the GT curves and the data are consistent with each other. Then, the GT curves show a dip. 
All of the GT curves, except for the last rapidity label $n=19$, have positive values. 
The final rapidity label $n=19$ curve shows a dip around $r\sim 0.5\,\text{fm}$.
In a log-log plot, this dip consists of three branches. The first and last branches have positive values, and the second branch has negative values.
In this case, there is clear disagreement between the JIMWLK evolution simulation data and the GT curves obtained from the adjoint dipole.

In conclusion, one should not use the GT to evaluate the TMD distributions, $\calf_{qg}^{(2)}$ and $\calf_{qg}^{(3)}$ without first correcting for the mismatch in the evolution speed observed for $\Omega_{qg}^\omega$.

\subsection{Gluon-Gluon sector}
Out of the seven TMD distributions considered in the gluon-gluon sector, we will illustrate only two of them, the reason being that the situation is similar to the quark-gluon sector.

\begin{figure}
    \centering
    \includegraphics[width=0.9\linewidth]{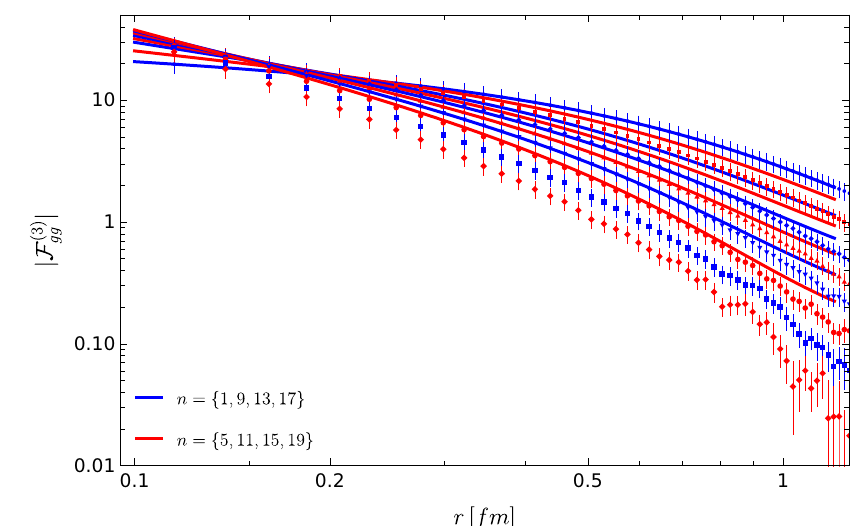}
    \hfill
    \includegraphics[width=0.9\linewidth]{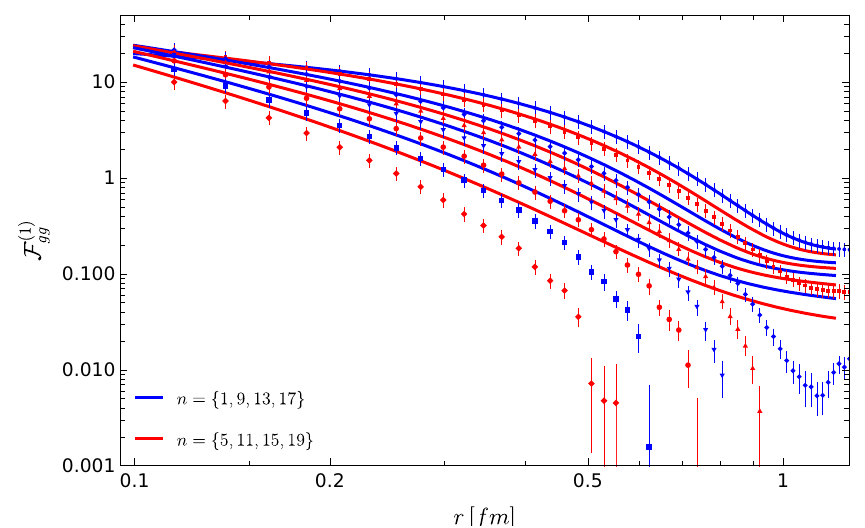}
    \caption{The TMD distributions $\calf_{gg}^{(3)}$ (top) and $\calf_{gg}^{(1)}$ (bottom).
    As the rapidity $n\, \alpha_s\delta Y$ increases, the distribution drifts toward the UV (colors are alternated between blue and red according to the figure labels).
    \PointsLinesAdj
    }
    \label{fig:GT_comp_GG}
\end{figure}

The top panel of {fig.~\ref{fig:GT_comp_GG}} shows the TMD distribution $\calf_{gg}^{(3)}$, which is the Weizs\"acker-Williams distribution, see the appendix~\ref{App:operators}.
It is directly proportional to $\Omega_{gg}^{1}$ and was discussed in the section {\ref{subsec:omega_gg}}.
Although the GT curve's functional form appears correct, the rapidity evolution is too slow compared to the simulation data.

The bottom panel of {fig.~\ref{fig:GT_comp_GG}} shows the TMD distribution $\calf_{gg}^{(1)}$.
Unlike the other plots in this manuscript, we have chosen to show the distribution rather than the absolute value. 
The first three sets of data $n=\{1,5,9\}$ (the rightmost data points) have positive values over the whole range in $r$.
The last five sets of data $n = \{11,13,15,17,19\}$ have positive values for small values of $r$, and become negative after a node located at $r_{\text{node},n} $. 
The corresponding nodes for the data $n = \{11,13,15,17,19\}$ are in the range $r_{\text{node},n} \in (0.5 \text{fm}, 1 \text{fm})$. 
Since the data are shown on a logarithmic scale, the negative branches are not shown at the bottom of {fig.~\ref{fig:GT_comp_GG}}. 
This illustration choice was made so that the bottom right corner of the plot would not be overloaded with the second branches of the distribution $n = \{11,13,15,17,19\}$ whose magnitude is of the order $\calo(10^{-2} - 10^{-1})$.
The simulation data shown in the bottom panel of {fig.~\ref{fig:GT_comp_GG}} illustrate the emergence of a node just after the rapidity label $n=9$ (blue data points).
This feature is completely absent from the GT curves, which have positive values across the entire rapidity range.

\subsection{Summary}

The conclusion follows the outcomes discussed in the quark-gluon sector: one should not evaluate the TMD distribution $\calf_{gg}$ using the GT without first correcting for the mismatch in the evolution speed observed for $\Omega_{gg}^\omega$.

\section{Discussion and prospects}
\label{sec:discussion}

In this study, we built upon previous research on small-$x$ TMD distributions by incorporating evolution into our framework in \cite{Cougoulic:2026drr}. 
We concentrated on the applicability of the Gaussian truncation, which is the preferred method for accessing these distributions due to its reduced computational complexity. 
Contrary to the results obtained for the MV initial condition, we found that, in the majority of cases, the Gaussian truncation does not correctly reproduce the results obtained from the full numerical solution of the JIMWLK equation. 
We demonstrated these conclusions using the expectation values of the $\hat{\Omega}_{ag}$ operators. Except for $\hat{\Omega}_{qg}^3$ and $\hat{\Omega}_{gg}^{8a,s}$, the Gaussian truncation was predicting a too slow evolution. 
As far as the TMD distributions are concerned, the discrepancies can take a more dramatic form because they are linear combinations of the $\hat{\Omega}_{ag}$ operators.
We showed, as examples, the $\calf_{gg}^{(1)}$ and $\calf_{qg}^{(3)}$ distributions, where additional nodes are either present or absent in the numerical data compared to the GT prediction. 
Our results for the $\hat{\Omega}_{ag}$ operators, and in particular the hierarchy of the speeds of the evolution of these distributions, may suggest a practical patch to the problem.
A possible rescaling of the rapidity variable depending on the involved dimension of the representation for each $\hat{\Omega}_{ag}$ operator could adapt the evolution speed of the GT to that of the full JIMWLK solution. 
However, a deeper understanding of the origin of the mismatch of speeds should be obtained first in order to better ground such a provisional solution.

To estimate the full phenomenological impact of these results, one must evaluate the Fourier transforms of the TMD distributions and combine them into the appropriate cross-sections. For example, the di-jets cross-section discussed in \cite{Marquet:2016cgx}. 
We leave that for a dedicated study because of the intricacies of the Fourier transforms of the TMD distributions. 
In addition to its connection to cross sections, evaluating TMD operators in momentum space is theoretically relevant for determining the speed at which TMD operators evolve and how this speed depends on $N_c$.
We expect this dependence to improve our understanding of the differences between, or interplay of, the BK and JIMWLK evolution equations.

The dipole gluon distribution has been considered beyond the eikonal limit in \cite{Li:2026azt}. 
By expanding the Lipatov vertex in eikonality, the eikonal contribution is given by $\calf_{qg}^{(1)}$, which is evaluated here.
The first correction to the Lipatov vertex was considered in \cite{Li:2026azt}, and it was shown that the resulting evolution equation sums single logarithms.
Extending our numerical framework to address the operators discussed in this analysis and evaluate their scale compared to that of the operator $\calf_{qg}^{(1)}$ evaluated here would be very interesting.
Other improvements to our analysis include incorporating the running coupling into the rapidity evolution, as well as considering its impact on predictions based on the Gaussian truncation. 
Extensions beyond the Gaussian truncation are also possible. Assuming \textit{time} locality for the evolution, it would be interesting to consider incorporating a three-point function in a manner similar to that employed in the MV model extensions, see {e.g} \cite{Jeon:2005cf}.

\acknowledgments
We would like to thank {T. Altinoluk}, {C. Marquet}, and {T. Stebel} for helpful comments on topics related to the evolution of TMD distributions and the Gaussian truncation.

\noindent

We gratefully acknowledge Polish high-performance computing infrastructure PLGrid (HPC Center: ACK Cyfronet AGH) for providing computer facilities and support within computational grants no. PLG/2023/016656, PLG/2024/017690, PLG/2026/019139.

We acknowledge Polish high-performance computing infrastructure PLGrid for awarding this project access to the LUMI supercomputer, owned by the EuroHPC Joint Undertaking, hosted by CSC (Finland) and the LUMI consortium through PLL/2025/08/018112 and PLL/2025/09/018947 grants.

P.K. and F.C. acknowledge support from the Polish National Science Center (NCN) grant No. 2022/46/E/ST2/00346.

\appendix
\section{Fit tables}\label{App:fit}
The fit of the initial condition shown in the left panels of both {fig.~\ref{fig:Fit_fun_dipole}} and {fig.~\ref{fig:Fit_adj_dipole}} with the functional form given in \req{eq:fit_model_IC} uses the following parameters:
\begin{equation}
    \mu_\text{fit} = 0.87; \qquad m_\text{fit} = 0.20.
\end{equation}
In the table \ref{tab:QG_GG}, we show the parameters extracted with the fit of \req{eq:fit_model_Evo} to the data.

\begin{table}[ht]
\begin{subtable}[c]{0.49\textwidth}
    \centering
    \begin{tabular}{|c|c|c|c|}
        \hline
         $n$ & $b_n$ & $c_n$ & $\delta_n$ \\
         \hline
         1 & -0.0073 & 0.018 & 0.84 \\
         2 & -0.0075 & 0.020 & 0.82 \\
         3 & -0.0074 & 0.021 & 0.78 \\
         4 & -0.0067 & 0.022 & 0.74 \\
         5 & -0.0060 & 0.024 & 0.71 \\
         6 & -0.0055 & 0.025 & 0.66 \\
         7 & -0.0051 & 0.027 & 0.64 \\
         8 & -0.0050 & 0.029 & 0.62 \\
         9 & -0.0054 & 0.031 & 0.60 \\
        10 & -0.0058 & 0.034 & 0.58 \\
        11 & -0.0043 & 0.037 & 0.56 \\
        12 & -0.0053 & 0.041 & 0.55 \\
        13 & -0.0051 & 0.045 & 0.54 \\
        14 & -0.0048 & 0.048 & 0.53 \\
        15 & -0.0056 & 0.053 & 0.52 \\
        16 & -0.0053 & 0.057 & 0.51 \\
        17 & -0.0069 & 0.062 & 0.50 \\
        18 & -0.0069 & 0.068 & 0.50 \\
        19 & -0.0076 & 0.073 & 0.49 \\
        \hline
    \end{tabular}
    \subcaption{Quark-gluon sector}
    \label{tab:QG}
\end{subtable}
\hfill
\begin{subtable}[c]{0.49\textwidth}
    \centering
    \begin{tabular}{|c|c|c|c|}
         \hline
         $n$ & $b_n$ & $c_n$ & $\delta_n$ \\
         \hline
         1 & -0.0088 & 0.020 & 0.88 \\
         2 & -0.0094 & 0.021 & 0.84 \\
         3 & -0.0098 & 0.022 & 0.80 \\
         4 & -0.0091 & 0.023 & 0.76 \\
         5 & -0.0084 & 0.024 & 0.72 \\
         6 & -0.0082 & 0.025 & 0.68 \\
         7 & -0.0079 & 0.027 & 0.65 \\
         8 & -0.0073 & 0.029 & 0.62 \\
         9 & -0.0079 & 0.031 & 0.61 \\
        10 & -0.0082 & 0.033 & 0.58 \\
        11 & -0.0066 & 0.036 & 0.56 \\
        12 & -0.0080 & 0.040 & 0.56 \\
        13 & -0.0074 & 0.044 & 0.54 \\
        14 & -0.0074 & 0.048 & 0.53 \\
        15 & -0.0090 & 0.052 & 0.53 \\
        16 & -0.0103 & 0.057 & 0.52 \\
        17 & -0.0103 & 0.062 & 0.51 \\
        18 & -0.0108 & 0.067 & 0.51 \\
        19 & -0.0135 & 0.074 & 0.51 \\
        \hline
    \end{tabular}
    \subcaption{Gluon-gluon sector}
    \label{tab:GG}
\end{subtable}
\caption{The parameters used to fit \req{eq:fit_model_Evo} to the data shown in {fig.~\ref{fig:Fit_fun_dipole}} and {fig.~\ref{fig:Fit_adj_dipole}}.}
\label{tab:QG_GG}
\end{table}

\section{Implementation of the JIMWLK evolution} 

The initial condition is generated according to the MV-model, and we refer to \cite{Cougoulic:2026drr} for further details. 
The evolution of WLs is explained in section \ref{subsec:JIMWLK}. The numerical implementation follows \cite{Rummukainen:2003ns,Lappi:2012vw,Cali:2021tsh}. The evolution equation in rapidity $y$ with a step of size $\delta y$ reads
\begin{multline}
\label{eq. main}
U(\mathbf{x},y+\delta y) = \exp\left( -i\sqrt{\delta y} \sum_{\mathbf{y}} U(\mathbf{y},y) \left(\mathbf{K}(\mathbf{x}-\mathbf{y}) \cdot \boldsymbol{\xi}(\mathbf{y},y) \right) U^{\dagger}(\mathbf{y},y) \right) \, U(\mathbf{x},y) \times \\ \times  
\exp \left( i\sqrt{\delta y} \sum_{\mathbf{y}} \mathbf{K}(\mathbf{x}-\mathbf{y}) \cdot \boldsymbol{\xi}(\mathbf{y},y) \right) \\
\equiv \exp(-i\sqrt{\delta y}B(\mathbf{x},y)) \, U(\mathbf{x},y) \, \exp(i\sqrt{\delta y}A(\mathbf{x},y)),
\end{multline}
where $\mathbf{K}(\mathbf{x})$ is a kernel function given by \ref{eq:evo_alpha} and $\boldsymbol{\xi}(\mathbf{x},y)$ are random vectors valued in the Lie algebra (with generators $t^a$) on each site of the lattice,
\begin{equation}
\boldsymbol{\xi}(\mathbf{x},y) = (\xi_x(\mathbf{x},y), \xi_y(\mathbf{x},y)) = (\xi_x^a(\mathbf{x},y) t^a, \xi_y^b(\mathbf{x},y) t^b) \,,
\end{equation}
from a normal distribution with unit width,
\begin{equation}
\label{eq. noise}
\langle \xi^a_i(\mathbf{x},y)\, \xi^b_j(\mathbf{y},y') \rangle = \delta^{ab} \delta_{ij}\, \delta(\mathbf{x} - \mathbf{y}) \, \delta(y - y'). 
\end{equation}
In order to use Fourier acceleration, we define 
\begin{equation}
\mathbf{\tilde{K}}(\mathbf{k}) = \frac{1}{\sqrt{L_x L_y}} \sum_{\mathbf{x}} e^{i \mathbf{k} \mathbf{x}} \mathbf{K}(\mathbf{x}) \qquad 
\boldsymbol{\tilde{\xi}}(\mathbf{k},y) = \frac{1}{\sqrt{L_x L_y}} \sum_{\mathbf{x}} e^{i \mathbf{k} \mathbf{x}} \boldsymbol{\xi}(\mathbf{x},y)
\end{equation}
Then
\begin{equation}
A(\mathbf{x},y) =  \frac{1}{\sqrt{L_x L_y}} \sum_{\mathbf{k}} e^{-i \mathbf{k} \mathbf{x}}   \tilde{\mathbf{K}}(\mathbf{k})  \tilde{\boldsymbol{\xi}}(\mathbf{k},y) \,.
\end{equation}
We also define a matrix $\mathbf{\mathcal{U}}$
\begin{equation}
\mathbf{\mathcal{U}}(\mathbf{y},y) = U(\mathbf{y},y) \boldsymbol{\xi}(\mathbf{y},y) U^{\dagger}(\mathbf{y},y)
\,,
\end{equation}
which we transform to momentum space
\begin{equation}
\tilde{\mathcal{U}}(\mathbf{k},y) = \frac{1}{\sqrt{L_x L_y}} \sum_{\mathbf{x}} e^{i \mathbf{k} \mathbf{x}}   \mathcal{U}(\mathbf{x},y) \,.
\end{equation}
Then, we have for $B$
\begin{equation}
B(\mathbf{x},y) = \frac{1}{\sqrt{L_x L_y}} \sum_{\mathbf{k}} e^{-i \mathbf{k} \mathbf{x}}  \tilde{\mathbf{\mathcal{U}}}(\mathbf{k},y)
\tilde{\mathbf{K}}(\mathbf{k})  \,.
\end{equation}

\section{Definition of the operators}\label{App:operators}
The finite difference symbol $D_i$ is given by
\begin{equation}
D_i V_\xvec \equiv \frac{1}{2} \left( V_{\xvec + e_i} - V_{\xvec - e_i} \right).
\end{equation}
where $e_i$ denotes the lattice step in the direction $i$.
Each derivatives of a WL appearing in the combination $\left( V^\dagger_\xvec \partial^i_\xvec V_\xvec \right)$ are replaced according to \cite{Marquet:2016cgx}:
\begin{equation}
\left( V^\dagger_\xvec \partial^i_\xvec V_\xvec \right) 
\longrightarrow
\mathcal{A}_\xvec^i = \frac{1}{2a} \left[ \left( V_\xvec^\dagger D_i V_\xvec \right) - \left( V_\xvec^\dagger D_i V_\xvec \right)^\dagger \right].
\end{equation}
where $a$ is the lattice spacing. This replacement is made to ensure anti-hermicity of the operator. We observe that the endpoints of the operator $\cala^i$ are both at $-\infty$. 
\begin{equation}
    \left( V^\dagger_\xvec \partial^i_\xvec V_\xvec \right)_{j\ell}  = \left(V_\xvec^\dagger[-\infty,+\infty]\right)_{jk} \left(\partial^i V_\xvec[+\infty,-\infty] \right)_{k\ell}
\end{equation}
This operator is thus conveniently used to express backward staples.
For convenience, we introduce the operator  $\mathcal{B}^i$ to be the reflection of $\cala^i$:
\begin{equation}
    \calb^i = (\partial^i V) V^\dagger = \left(\partial^i V[+\infty,-\infty]\right) V^\dagger[-\infty,+\infty],
\end{equation}
which has both endpoints at $+\infty$. This operator will be convenient to express forward staples.
To enforce the operators $\cala^i$ and $\calb^i$ to be in the adjoint irrep, we also introduce the corresponding projection:
\begin{subequations}
\begin{align}
\cala^i &\longrightarrow 2\, \tr \left[ \cala^i t^a\right] t^a = \cala^i - \frac{1}{N_c} \tr \left( \cala^i \right) \mathbb{1}\\
\calb^i &\longrightarrow 2\, \tr \left[ \calb^i t^a\right] t^a = \calb^i - \frac{1}{N_c} \tr \left( \calb^i \right) \mathbb{1}
\end{align}
\end{subequations}
where we used Fierz identity on the {r.h.s.} of the equal sign. This will ensure that traces of $\cala$ and $\calb$ vanish for any discretization length $a = L/N$ used for the numerical evaluations.

\paragraph{Quark - Gluon:} according to \cite{Marquet:2016cgx,Bury:2018kvg}, we write TMD operators as:
\begin{subequations}\label{def_TMDqg_WL}
\begin{align}
    \hcalf_{qg}^{(1)}(\xvec,\yvec) 
    &= \tr \left[ (\calb^i_\xvec V_\xvec)^\dagger \calb^i_\yvec V_\yvec \right],\\
    -N_c \hcalf_{qg}^{(2)}(\xvec,\yvec) 
    &= \tr\left[\calb^i_\xvec \calb^i_\yvec\right] \tr \left[ V_\yvec V^\dagger_\xvec \right]\\
    -\hcalf_{qg}^{(3)}(\xvec,\yvec) 
    &= \tr \left[ \calb^i_\xvec V_\yvec V^\dagger_\xvec \calb^i_\yvec \right]
\end{align}
\end{subequations}

\paragraph{Gluon - Gluon:} according to \cite{Marquet:2016cgx,Bury:2018kvg}, we write TMD operators as:
\begin{subequations}\label{def_TMDgg_WL}
\begin{align}
    N_c \hcalf_{gg}^{(1)}(\xvec,\yvec)
    &= - \tr \left[V_\xvec^\dagger \calb^i_\xvec \calb^i_\yvec V_\yvec\right] \tr \left[ V_\xvec V^\dagger_\yvec\right] 
    = \hcalf_{qg}^{(1)} \ \tr \left[ V_\xvec V^\dagger_\yvec\right], \\
    -N_c \hcalf_{gg}^{(2)}(\xvec,\yvec)
    &= \tr \left[ \calb^i_\yvec V_\yvec V_\xvec^\dagger\right]  \tr \left[ \calb^i_\xvec V_\xvec V^\dagger_\yvec\right],  \\
    - \hcalf_{gg}^{(3)}(\xvec,\yvec)
    &= \tr \left[ \calb^i_\xvec \calb^i_\yvec\right], \\
    - \hcalf_{gg}^{(4)}(\xvec,\yvec)
    &= \tr \left[ \cala^i_\xvec \cala^i_\yvec\right], \\
    -\hcalf_{gg}^{(5)}(\xvec,\yvec)
    &= \tr \left[ V_\yvec V^\dagger_\xvec \calb^i_\yvec V_\xvec V_\yvec^\dagger \calb^i_\xvec \right] \\
    -N_c^2 \hcalf_{gg}^{(6)}(\xvec,\yvec)
    &= \tr \left[ \calb^i_\xvec \calb^i_\yvec\right]  \tr \left[V_\xvec V^\dagger_\yvec\right]  \tr \left[ V_\yvec V_\xvec^\dagger \right] ,\\
    -N_c \hcalf_{gg}^{(7)}(\xvec,\yvec)
    &= \tr \left[ V_\xvec V^\dagger_\yvec  \calb_\xvec^i \calb^i_\yvec\right] \tr\left[V_\yvec V_\xvec^\dagger \right].
\end{align}
\end{subequations}

\noindent
Further details can be found in our previous analysis on the initial condition of those TMD distributions \cite{Cougoulic:2026drr}.

\bibliographystyle{ieeetr}
\bibliography{draft_bib}
\end{document}